\documentclass[12pt,preprint]{aastex}

\newcommand{\ie}{i.e.\ }
\newcommand{\be}{\begin{equation}}
\newcommand{\ee}{\end{equation}}
\newcommand{\beq}{\begin{eqnarray}}
\newcommand{\eeq}{\end{eqnarray}}

\newcommand{\vect}[1]{{\bf #1}}

\usepackage{amsthm}
\usepackage{amsmath}
\usepackage{amsfonts}
\usepackage{graphicx}
\usepackage{times}

\begin{document}
\title{Comparison of seismic signatures of flares obtained by SOHO/MDI and GONG instruments}

\author{S. Zharkov\altaffilmark{1}, V.V.Zharkova\altaffilmark{2}, S.A. Matthews\altaffilmark{1} }

\altaffiltext{1}{Mullard Space Science Laboratory,
          University College London, Holmbury St. Mary, Dorking, RH5 6NT, 
          UK}
\altaffiltext{2}{Horton D building, 
	Department of Mathematics,
	University of Bradford,
          Bradford, BD7 1DP,
          UK}

\begin{abstract}
 The first observations of seismic responses to solar flares were carried out using time-distance (TD) and holography techniques applied to SOHO/MDI Dopplergrams obtained from space and un-affected by terrestrial atmospheric disturbances. However, the ground-based network GONG is potentially a very valuable source of sunquake observations, especially in cases where space observations are unavailable. In this paper we present updated technique for pre-processing of  GONG observations for application of subjacent vantage holography. Using this method and TD diagrams we investigate several sunquakes observed in association with M and X-class solar flares and compare the outcomes with those reported earlier using MDI data. In both GONG and MDI datasets, for the first time, we also detect the TD ridge associated with the September 9, 2001 flare. Our results show reassuringly positive identification of sunquakes from GONG data that can provide further information about the physics of seismic processes associated with solar flares.

\end{abstract}


\section{Introduction}

Discovery of sunquakes associated with solar flares 
\citep{kz1998,Donea1999} has opened a new era in the  investigation of energy and momentum transport mechanisms 
from the upper atmosphere to the photosphere and beneath, uncovering structure of these spectacular events.
Sunquakes, seen as circular or elliptical waves - ripples, propagating outward from impulsive hard X-ray (HXR) solar flares along the solar surface, appear 20-60 minutes after the flare onsets. 
The surface ripples are also associated with strong downward shocks preceding these ripples with close (1-4 minutes) temporal correlation with the start of HXR flares, indicating that high energy particles play some role in initiation of sunquakes \citep{Zharkova07, Martinez2008a}. 

Even though every flare is expected to inject particle beams of one kind or another into a flaring atmosphere, inducing either shocks or magnetic impulses, not many of them have { recorded measurable} signatures of seismic activity. Initially only X-class flares were considered as candidates for producing sunquakes. 
The first flare detection used time-distance diagrams and reported well distinguished ripples emanating from the center of the location of a hard X-ray source in the X1.1 flare 9 July 1996 \citep{kz1998, Donea1999}. It was followed by detection of quakes associated with two extremely powerful solar flares of class X17 and X10 which erupted in NOAA Active region 10486 on October 28 and 29, 2003. These two flares, known as the Halloween 2003 flares, were extensively investigated in \cite{DL2005} by applying subjacent vantage acoustic holography to SOHO/MDI Dopplergram data. The  acoustic signatures were also shown to co-align with the hard X-ray signatures and GONG intensity observations revealed significant radiative emission with a sudden onset in the compact region encompassing the acoustic signature. 

Further analysis of the 29 October 2003 quake was presented in \citet{LD2008}, where the authors proposed a new method for correcting { intensity data recorded by the Global Oscillation Network Group (GONG)}, that allowed comparison of acoustic kernels with white-light traces of the flare. Later \citet{VZharkova2006, K2006, Zharkova07} investigated the Halloween flare of 28 October 2003 using the time-distance diagram method applied to SOHO/MDI data and detected three distinct seismic sources that coincided with the holographic sources from \cite{DL2005}. { For the 29 October 2003 flare there were no time-distance ridges found by the authors of \citet{Zharkova07} when they analyzed MDI dopplergrams, however one was later reported in \cite{K2006}}. The first M-class flare in which seismic signatures were detected  by means of acoustic holography using Doppler velocity data from SOHO/MDI instrument was the flare that occurred in NOAA Active Region 9608 on September 9, 2001 \citep{Donea2006}.  

Later the list of acoustically active flares was significantly extended. 
 \citet{BI05} reported another six such flares, adding another four in \citet{DoneaEtAl2006}.
In fact, during the period of observations with SOHO/MDI instrument there have been 17 flares showing signs of acoustic activity possibly related to sunquakes\footnote{see \url{http://users.monash.edu.au/$\sim$dionescu/sunquakes/sunquakes.html}}.
This expanding list motivated researchers to look further and to explore the data from the  ground-based GONG observatories, which offers better coverage of helioseismic data compared to  SOHO/MDI. 

In the unusually quiet solar minimum between cycles 23 and 24 more attention was paid to each new flare occurring on the Sun, one of which was the flare of 14 December 2006. { The latter was observed only by the GONG instruments and revealed some noticeable seismic signatures} in both time-distance ridges and egression powers \citep{Matthews10}.   In order to validate { these findings, a comparison is required of the signatures of sunquakes derived for both the time-distance and holographic techniques} from GONG data with those from SOHO/MDI for a number flares with distinct seismic signatures. Such a comparison will allow us  { to understand the differences in appearances and} to derive recommendations for the reliable detection of sunquakes from GONG data.

In this study we consider three acoustically active flares that have the luxury of  helioseismic observations available from both the GONG and the SOHO/MDI instrument. { The available Dopplergrams (GONG and MDI) are used to  analyze the acoustic signatures of the flares by using both the time-distance diagram technique \citep[TD method;][]{kz1998} and acoustic holography \citep[e.g.][]{BL1999, LB2000}.} The description of data and additional corrections for the technique applied to GONG data are presented in section \ref{descrip}, the results of the comparison are described in section \ref{valid} and conclusions with recommendations are drawn in section \ref{conc}. 

\section{Description of data and techniques} \label{descrip}
In this study we use three acoustically active flares with strong seismic signals detected by SOHO/MDI. The first flare is an M-class flare that occurred in NOAA Active Region 9608 on September 9, 2001. The solar quake associated with the flare was the first one detected { for M-class flares} and investigated by means of acoustic holography in \citet{Donea2006} using data from SOHO MDI. The other two are extremely powerful X-class flares that erupted in NOAA Active region 10486 on October 28 and 29, 2003, with associated sunquakes first detected using acoustic holography by \cite{DL2005}. The Halloween flares have also been investigated by applying time-distance method to SOHO/MDI data in \citet{Zharkova07}, where three distinct sources were detected for October 28 flare. However, after extensively analyzing MDI velocity data for the 29 October 2003 flare,  the authors of \citet{Zharkova07} did not find a time-distance ridge in any of time-distance diagrams computed around the flare location. 

\subsection{Helioseismic techniques for quake detection} \label{techn}

In order to detect and analyze the solar quake associated with a flare we use both time-distance analysis 
\citep{kz1998} and acoustic holography \citep{Donea1999}. Time-distance analysis is applied to detect the circular ripples
generated by the quake. This consists of rewriting the observed surface signal in polar coordinates relative to the source, i.e. $ v(r, \theta, t)$,  
and using azimuthal transformation 
\be
V_m(r, t)=\int_0^{2\pi} v(r, \theta, t) e^{-i m \theta} d\theta,
\label{equ:td}
\ee
to study the $m=0$ component for evidence of the propagating wave. Then if seen, the quake manifests itself 
as a time-distance ridge, thus providing estimates of the surface propagation speed and the time of 
excitation. In this work the GONG high-cadence velocity data were used in the time-distance analysis.

Acoustic holography is applied to calculate the egression power maps from observations. The holography method 
\citep{BL1999, Donea1999, BL2000a, LB2000} works by essentially ``backtracking'' the observed surface 
signal, $ \psi(\vect{r}, t)$, by using Green's function, $G_+ (|\vect{r}-\vect{r}'|, t-t') $, which 
prescribes the acoustic wave propagation from a point source. 
{ This allows us to reconstruct egression 
images showing the subsurface acoustic sources and sinks. Following \citet{DL2005}, in temporal Fourier domain we have
\be
\hat H_+({\vect{r}, \nu})= \int_{a<|\vect{r}-\vect{r}'|<b} d^2\vect{r}' \hat G_+ (|\vect{r}-\vect{r}'|, \nu) \hat \psi(\vect{r}', \nu),
\label{equ:holo0}
\ee
where $a, b$ define the holographic pupil and $\hat H_+({\vect{r}, \nu})$ is the temporal Fourier transform of $H_+({\vect{r}, t}).$ Then
\be
H_+({\vect{r}, t})= \int_{\Delta \nu} d\nu \, e^{2\pi i \nu t} \, \hat H_+({\vect{r}, \nu}),
\label{equ:holo1}
\ee
whence the square amplitude of egression  is called the egression power
\be
P( \vect{r}, t)=| H_+({\vect{r}, t}) |^2 dt .
\label{equ:holo2}
\ee
Green's functions built using a geometrical-optics approach are used in this study.
As flare acoustic signatures can be submerged by ambient noise for the relatively  long periods over which the egression power maps are integrated, again we follow \citet{DL2005} using egression-power 'snapshots' to discriminate flare emission from the noise with pass-band integration in equation (\ref{equ:holo1}) performed over positive frequencies only in order to reduce noise. Such a snapshot is simply a sample of the egression power within a time $\Delta t=\frac{1}{2 {\rm \  mHz}}=500 {\rm s}$. 
The snapshots used in this work are taken from the egression power, $P(\vect{r}, t)$, at selected times, $t$.

\subsection{Observations and data reduction}

An M9.5-class flare occurred in NOAA Active Region 9608 around 20:40 UT on September 9, 2001 at around $104^\circ$ Carrington longitude and $26^\circ$ latitude south. GOES soft X-ray flux reached a peak at 20:46 UT, with the background emission remaining above the ''C'' level for most of the flare period. The flare of October 28, 2003, occurred in NOAA Active Region 10486 at $287^\circ$ Carrington longitude and $8^\circ$ south latitude. It is classified as X17.2, one of the most powerful amongst the quake producing flares recorded. The GOES satellite detected increased X-ray flux starting at 09:51 UT, reaching a maximum at 11:10 UT. On the following day in the same Active Region an X10 class flare occurred at around $270^\circ$ Carrington longitude, $10^\circ$ latitude in the southern hemisphere. The X-ray flux observed by GOES began to increase at 20:41 UT, reaching maximum at 20:49 UT and ending at 21:01 UT.

The helioseismic observations analyzed in this study were obtained by  the GONG \citep{Harvey1988, GONG1996} ground-based observatories, and by the MDI instrument \citep{scherrer1995} onboard the SOHO spacecraft. Both GONG and MDI observe using the photospheric Ni I 6768 \AA\ line. GONG  observations are made with one minute cadence and normally include full-disk Dopplergrams, line-of-sight magnetograms and intensity images. In this work we use MDI   full-disk Doppler images obtained with a one minute cadence. Velocity measurements in both cases are made from Doppler shifts of the Ni spectral line. MDI estimates the velocities from line instensities (filtergrams) scanned in several locations across the line, while GONG relies on a fast Fourier tachometric scans across the line \citep{Harvey1988} to derive the surface-velocity images based on a standard response of the line profile to Doppler motion caused by propagation of acoustic waves.

For the October 28, 2003 flare we use two-hour-long full-disk velocity observations with one minute cadence from the SOHO/MDI instrument and GONG starting at 10:46 UT. The October 29, 2003 and September 9, 2001 series commence at 20.00 UT on the corresponding dates. In addition, for the Halloween flares we use one-minute cadence intensity observations available from the GONG for the same period. Unfortunately, there are no such intensity  observations available for the duration of the September 9 flare. Following the standard approach in local helioseismology, we extract datacubes centered on the region of interest from each full-disk series to remap the data onto the heliographic grid using Postel-projection and to remove a differential rotation at the Snodgrass rate. For the velocity data the series average full-disk velocity image is subtracted from each observations before the procedure, in order to remove the rotation gradient. Due to different resolution of the instruments, SOHO MDI data is remapped at 0.125 degrees per pixel resolution, while GONG datacubes are at 0.15 degrees per pixel. 

For acoustic holography we  use Green's functions centered at 6mHz, so the datacubes  are filtered in the frequency domain using a bandpass filter allowing  the full signal in the 5-7 mHz band, with steep Gaussian roll-offs on each side. The pupil dimensions for each dataset for the selected flares are presented in Table \ref{tab:pupil}.

\subsection{Additional corrections for  the GONG data}
\label{sec:addGONG}
As GONG is a ground-based network, its data are affected by visibility conditions at the time of observation. 
{Effects such as atmospheric smearing and local stochastic translation introduce spurious temporal variations in magnetized regions that can easily dominate over a genuine seismic signal.  Another concern can be related to GONG usage of tachometric scanning of the Ni  line, which, due to variations in atmospheric conditions between the start and end of the scan, is likely to be more affected \citep[see, for example,][]{GK1988}.  

The spurious Doppler shifts cited by \citet{GK1988} are  applicable to radiation incidence away from normal incidence (above 2$^\circ$) passing through a Fabrey-Perot etalon, which has effective path differences of $2.2\times10^4$ wavelengths implemented for GONG  \citep{HarveyGONG1995}.  
The GONG optics ingeniously avoid this problem by directing the long optical path through glass and the shorter through air, the geometrical paths being the same to within about a micron \citep{TitleRamsey1980, HarveyGONG1995}.
}

In the presence of a strong magnetic field (e.g. sunspot umbrae) Zeeman splitting of a magnetic line introduces spurious phase shifts in the measurements \citep{Rajaguru2007}. 
 One possible reason for such effect is the reduced line intensities within a sunspot  \citep{TonerLabonte1993, Bruls1993, Norton2006} which cause noise such as due to variable atmospheric smearing to introduce spurious intensity observations
 from surrounding region into the desired pixel \citep{TonerLabonte1993,Braun1997}.

This, for instance, leads to significant differences in the computed acoustic power maps between the sunspot data from MDI and GONG, with the space-based data generally showing suppression of the acoustic power over a sunspot region  \citep[e.g.][and references therein]{Gizon2009}, while the ground-based GONG data demonstrating a large power increase at the same location (see top row of Figure \ref{fig:20010909_egression} for example) that is clearly noise-related due to the reasoning above.

To correct the atmospheric contribution in GONG observations in the first instance we use the method developed in \citet{LD2008} where the intensity data are available, e.g. Halloween flares. The method works by measuring atmospheric seeing effects such as translation and smearing of GONG intensity observations in relation to a reference image and then removing their contribution from the intensity data. As both intensity and velocity data come from the same instrument, we apply the parameters extracted from the intensity series to correct the line of sight velocity data.

In addition, since the atmospheric noise affects mostly the measurements taken over magnetized regions, we seek to minimize the contribution of such data to the quake-specific computation of egression power. First, we note that in the magnetized regions the atmospheric noise manifests itself as spurious velocity fluctuations leading to a substantial increase of the observed acoustic power. Such an increase in GONG data is assumed to be induced by the atmospheric seeing effects. Second, it is known that flares and associated with them solar quake sources are usually located over or near a sunspot. On the other hand, quake signatures such as ripples and time-distance ridges are normally seen in the surrounding non-magnetised region. This is, at least in part,  due to the complex and less understood propagation of magneto-acoustic waves generated by the quake in the sunspot itself. { Also, for our estimates of the egression power we use Green's function based on a non-magnetic model of the solar interior as it is intrinsic to the acoustic holography. In the view of equations (\ref{equ:holo1}-\ref{equ:holo2}) it is then reasonable to minimize the contribution of the velocity data taken over magnetic regions (MRs) such as sunspot.}

For smaller sunspots this can be achieved by the choice of pupil, ensuring that the smallest radius is always selected outside of MR. 
When a sunspot is large, other methods will need to be considered such as weighting of the sunspot data in the velocity measurements. One possible option is to fully neglect such data, \ie using zero as weights for sunspot velocities. This, of course, introduces artificial inhomogeneities in the computed egression power, but the qualitative strength of the quake source can still be evaluated by comparison with the egression power of surrounding plasma. In our experience, however, the best results are achieved by weighting all measurements by the inverse averaged acoustic power computed for the filtered velocity series. This is equivalent to normalizing the acoustic power of  the filtered datacube, similar to the approach of \citet{Raj2006} developed for time-distance helioseismology and phase-speed filtering. 
All of the GONG data used in the following sections are processed as described unless otherwise stated.

\begin{table}
\begin{center}
	\begin{tabular}{| c | c | c |}
	\hline
	\hline
	Flare &  MDI pupil size &  GONG pupil size \\
	\hline
	September 9, 2001 & 15-60 Mm	 & 25-70 Mm \\
	October 28, 2003  & 15-45 Mm & 25-95 Mm \\ 
	October 29, 2003 & 15-45 Mm & 20-55 Mm \\	
	\hline
	\end{tabular}
\end{center}
\begin{center}
	\caption{Holography pupil sizes for each dataset  for the selected flares. \label{tab:pupil}}
\end{center}
\end{table}

\section{Comparison of the GONG and MDI results}\label{valid}

\subsection{Comparison of acoustic holography results}
For illustration purposes, the results of calculations of the total egression power in 5-7 mHz range estimated from the MDI and GONG datacubes corresponding to observations of the September 9, 2001 flare, are presented in the bottom row of Figure \ref{fig:20010909_egression}. One can see that even after the corrections, the acoustic source suppression over the sunspot region is considerably weaker in the map computed from GONG data. This is generally the case for other observations we have considered and  is believed to occur because of the lower resolution and atmospheric noise contamination in the ground-based network's data. For these reasons, in order to reduce such contamination we consider larger pupil sizes when working with GONG data as shown in Table \ref{tab:pupil}.  By choosing the larger lower limit on pupil dimensions we ensure the minimised contribution of the measurements taken over MR for egression power computation at the points near and around sunspots. 

\subsubsection{Holography: September 9, 2001} \label{hol_sept}
Egression power snapshots computed for  the September 9 flare are presented in Figure \ref{fig:20010909_holo}, with the MDI data plotted in the left column and the GONG data on the right.  Velocity images averaged over the series duration for both instruments (located at the top of the  Figure \ref{fig:20010909_holo}) demonstrate clearly the differences in the original datasets, which are due to a further loss of resolution due to the atmospheric effects as described by \citet{LD2008}. Our MDI egression measurements for this flare essentially duplicate the results of \citet{Donea2006}. Comparing these with the obtained GONG snapshots (Figure \ref{fig:20010909_holo}), it is clear that even after the corrections, though many similarities are present, there is a significant variation between the two sets of images.

The most obvious difference is the apparent absence of the region with low acoustic emission around the sunspot in the GONG produced data.   Such an absence is clearly related to a much weaker signature of this phenomena in the GONG egression power seen in Figure \ref{fig:20010909_egression}. This  can be explained by the spurious atmospheric noise affecting GONG measurements over the regions with strong magnetic field. Nonetheless, the quake's signature is clearly present in the GONG measurements, with the locations of acoustic kernels agreeing very well for the two instruments. 
{
We note, however, the difference in acoustic kernel shape. This is, most likely, due to the reduced spatial resolution of GONG data suppressing higher-$l$ contribution to the egression. The possibility of atmospheric noise contamination is discussed later in Section \ref{conc}.
}

\subsubsection{Holography: October 28, 2003} \label{hol_28}
Egression power snapshots computed for the October 28 flare are presented in Figure \ref{fig:20031028_holo} with the reference GONG intensity image located in the top left corner, followed to the right by the MDI egression power snapshot taken at 11:07 UT with the arrows pointing to the detected acoustic kernels. This image essentially duplicates the panel c) of Figure 3 in \citet{DL2005}. For a reference, the GONG snapshot for the same time is presented in the bottom row with and without arrows. Here, one can see the acoustic signatures at the same locations as in the MDI data. Again, we note the region with weaker lower acoustic emission as seen by GONG, which affects the visible contrast of the quake kernels relative to their surroundings. It is also clear that while the locations are the same, the shape and strength of each of the four kernels varies from one instrument to another. For example, source 1 (see Figure  \ref{fig:20031028_holo}) appears to be more prominent and extended in GONG measurements compared with MDI. Given the reservations about ground-based data outlined above with the fact that our correction procedure rather artificially modifies the oscillation amplitudes in GONG data, it is clear that MDI measurements are closer to the true picture of the event. Nonetheless, we reiterate that quake signatures are clearly visible in GONG measurements in the same locations as those detected by MDI.

\subsubsection{Holography: October 29, 2003}
Egression power snapshots computed for the October 29 flare are presented in Figure \ref{fig:20031029_holo}  with the reference GONG intensity image located in the top left corner, followed to the right by the MDI egression power snapshot at 20:43 UT with the arrows pointing to the detected acoustic kernels. This image is essentially equivalent to the middle panel of Figure 6 in \citet{DL2005}. The corresponding GONG egression snapshot with the quake signature is plotted in the lower row on the left clearly present at approximately the same location as in the MDI plot. As an example, the egression power computed from the GONG velocity observation with the masked sunspot area is plotted in the bottom row to the right.

\subsection{Time-Distance diagrams}
\subsubsection{9 September 2001 flare} \label{td_sep}
The time-distance diagrams extracted from the MDI and GONG data are presented in Figure \ref{fig:20010909_TD}.
The ridge representing the quake in the MDI image is relatively weak but can be clearly seen. As far as we know, this is the first time-distance ridge for a solar quake associated with an M-class flare has been found. By comparing the MDI image with GONG time-distance diagram one can also detect in GONG a very similar disturbance located at the same part of the image.  Although weaker and less defined than in MDI, nevertheless, the ridge is definitely present.  Once again the relative weakness of the ridge can be explained by the fact that it is  obscured by a significant noise contribution. 
As additional re-assurance, there is a near perfect coincidence between the  MDI and GONG time-distance source locations. Also, as can be seen from Figure \ref{fig:20010909_holo} where the location of time distance source is marked as plus sign on selected GONG plots, there is a good agreement between the egression  acoustic kernels and time-distance source locations. 

\subsubsection{Halloween flares}
We were not able to find any distinguishable time-distance ridges for the October 29 flare in either MDI or GONG data, similar to the previous attempts \citep[e.g.][]{Zharkova07}.
The other GONG dataset for the October 28 flare, has a gap of about ten minutes between 11:30 and 11:40 UT. However, we have attempted to build the time-distance diagram and the time-distance plot obtained from the interrupted GONG data. This is presented in Figure \ref{fig:20031028_TD}. It shows (at least a part of) a ridge, with the location corresponding to Source 1 in \cite{Zharkova07}, reproduced in the top row of  Figure \ref{fig:20031028_TD}. { Once again, the location of the time distance ridge coincides with that of MDI and agrees with the egression measurements presented in section \ref{hol_28}}.

\section{General discussion and conclusions} \label{conc}
In this study we have compared egression power maps and time-distance diagrams derived from GONG and MDI data. SOHO MDI and GONG velocity datasets were used for three flares: M-class September 9, 2001 (Figures \ref{fig:20010909_egression}-\ref{fig:20010909_holo}, \ref{fig:20010909_TD}), X-class October 28, 2003 (Figures \ref{fig:20031028_holo} and \ref{fig:20031028_TD}), X-class October 29, 2003 (Figure \ref{fig:20031029_holo}). 

Reassuringly, the egression power map snapshots show seismic signatures common to both instruments for all flares. These signatures display an excellent agreement between the two instruments in terms of their time and location. We note, however, that even after the pre-processing, as outlined in Section \ref{sec:addGONG}, GONG egression measurements remain relatively noisy. This leads to important differences from the MDI produced egression maps, which are only partially compensated by increasing the pupil sizes when working with GONG data. One such difference is the apparent variance in shape and strength of the detected acoustic kernels as seen by these two instruments. Another is the relative weakness of the suppression of acoustic sources below the sunspot. Since solar quakes are often located in the sunspot, this means that identifying such seismic signatures in GONG egression measurements is a harder task due to its lower signal to noise ratio over the sunspot region. 

One such method of verification is the computation of the time-distance diagram. As we have demonstrated for September 9, 2001 and October 28, 2003 flares, such diagrams computed from GONG velocity data can present the additional evidence of the quake. Figures \ref{fig:20010909_TD} and \ref{fig:20031028_TD} clearly show that, in spite of being less sensitive, the GONG data can respond to the time-distance analysis producing noticeable ridges similar to those observed from the higher-quality MDI data. 
Results of the comparison with MDI time-distance measurements have again revealed an excellent spatial agreement between the two instruments in terms of the time-distance source location. 
Additionally, the fact that the locations of the sources observed with the GONG time-distance diagrams coincide with the acoustic kernels deduced from the GONG egression snapshots confirms that with these different techniques one observes the same events - seismic signatures induced by solar flares.
Therefore, we conclude that the GONG data can respond to time-distance analysis. Obviously, due to the characteristic noise, not every quake can be expected to produce the time-distance ridge in GONG diagrams, but  if a ridge is seen in the GONG ground-based data, one can expect that it will also be observed by using the higher-quality satellite MDI or HMI data.

We believe, the results of this study show that quake detection based on helioseismic reduction of GONG observations is possible. 
However, as the data are subject to atmospheric smearing and other related instrumental effects, GONG observations clearly have less intrinsic sensitivity than the space-borne observations. 
A useful prospective object for further study might be the quantitative comparison of intrinsic sensitivities of ground- and space-based helioseismic observations under various seeing conditions.
Nonetheless these results should allow us to add to the list of known sun-quakes by investigating the known flares in the Solar Cycle 23 using GONG data when MDI observations were not available. This will provide further information about the physics of seismic processes associated with solar flares.

\acknowledgments
\section{Acknowledgments}
The authors would like to thank Dr Charlie Lindsey for many useful discussions providing the stimulus to complete this work.

\begin{figure}
\centering
\includegraphics[width=8.2cm]{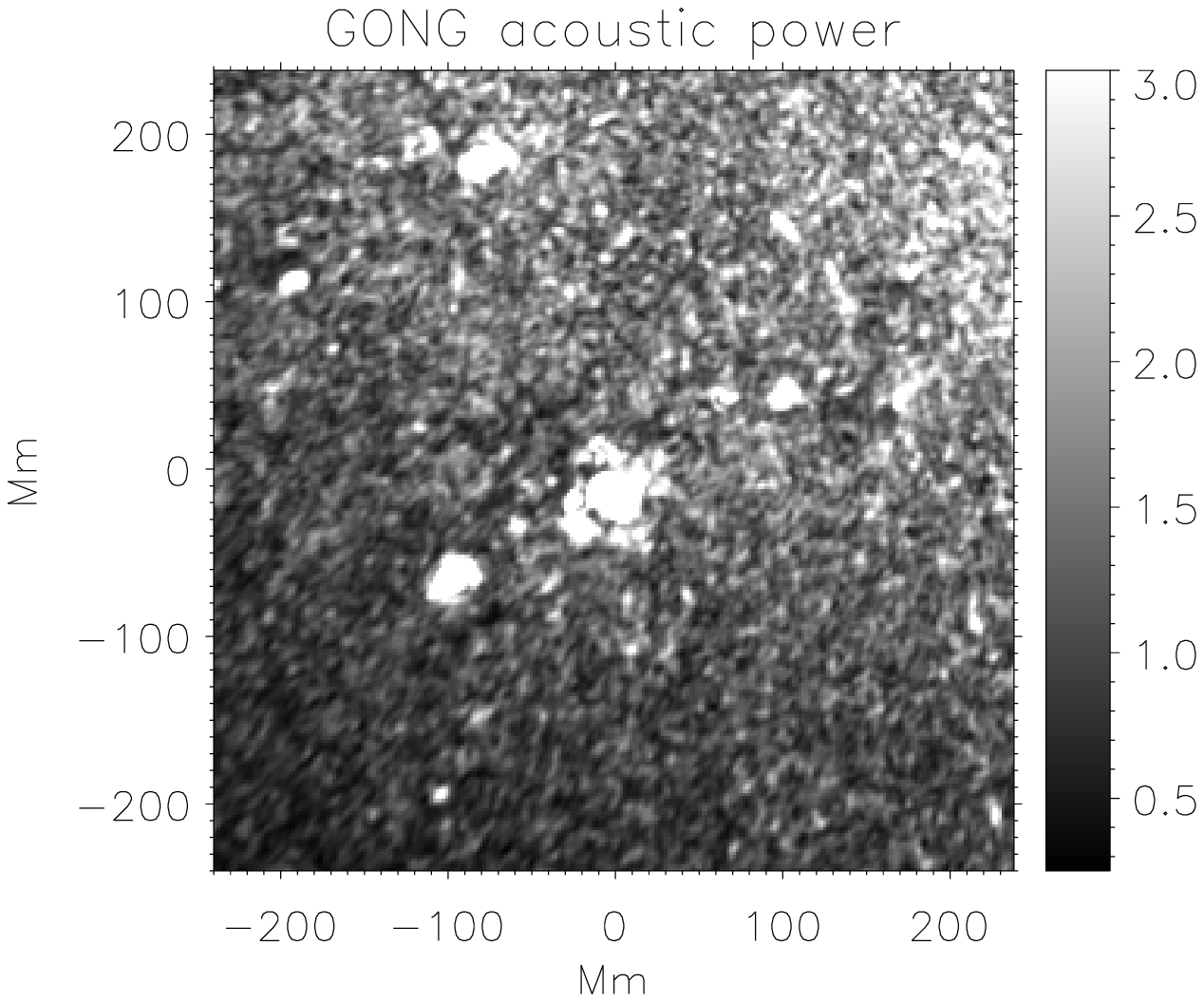} 
\includegraphics[width=8.2cm]{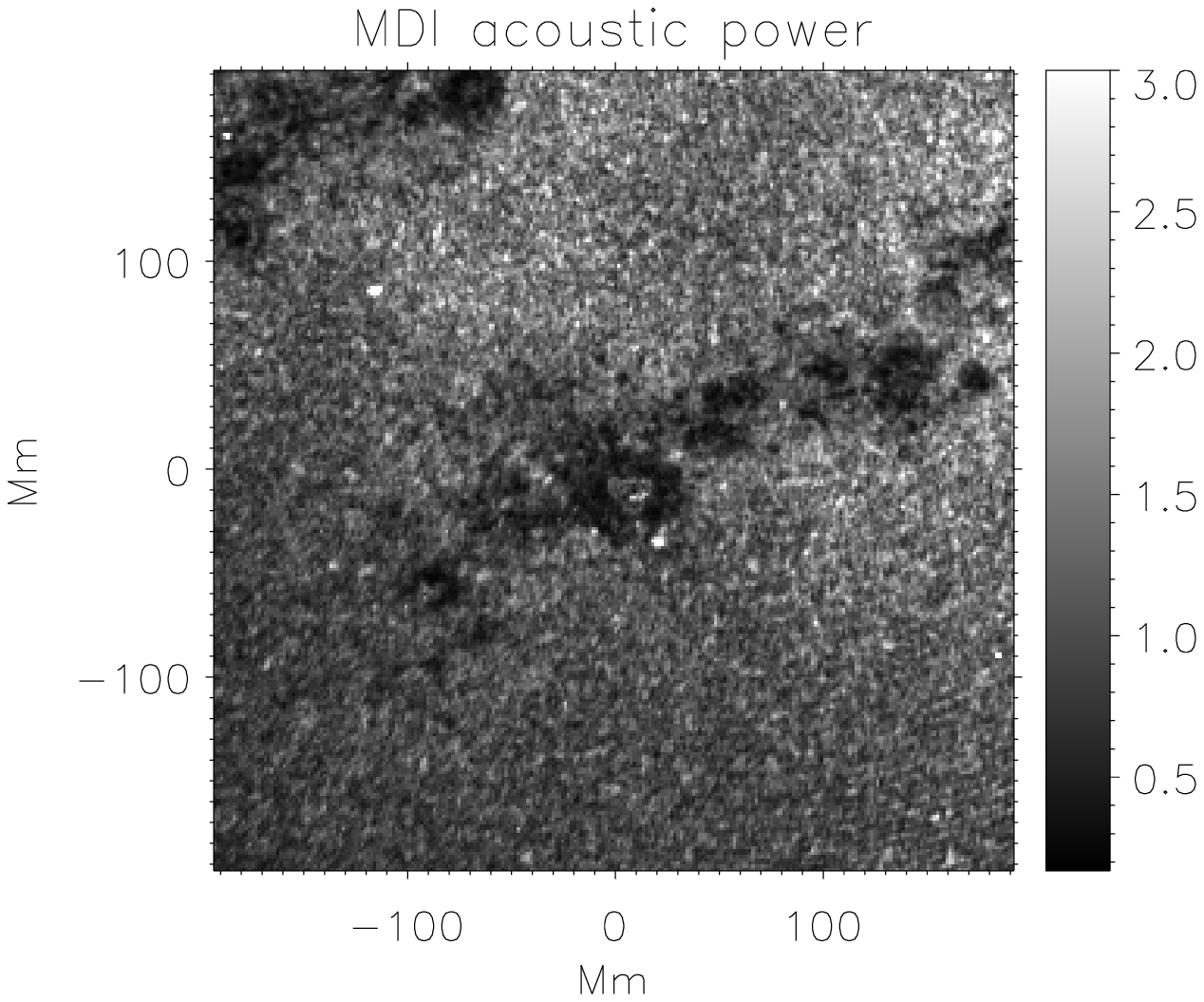} 
\\
\includegraphics[width=8.2cm]{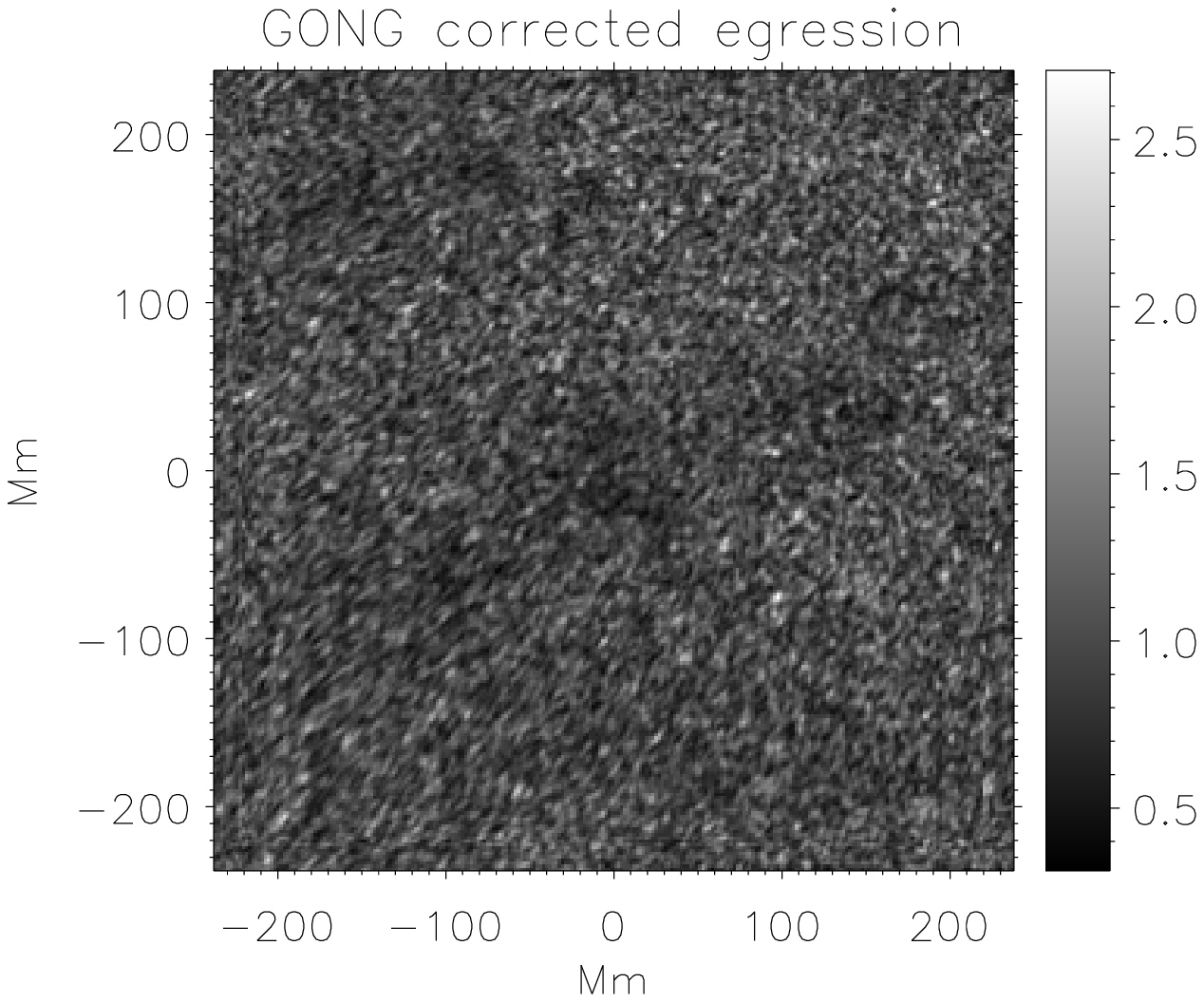} 
\includegraphics[width=8.2cm]{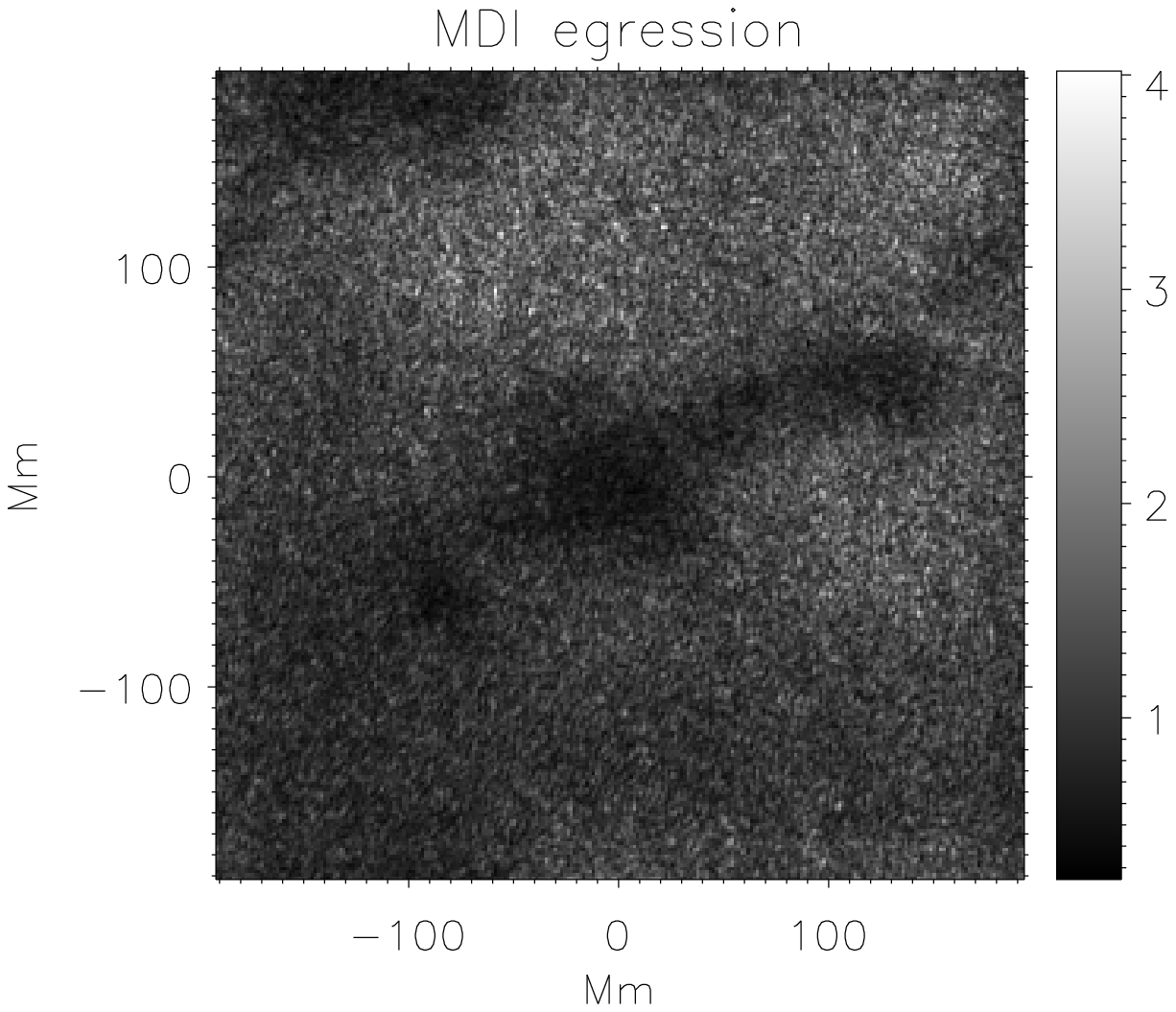} 
\caption{ September 9, 2001 M-class flare:  comparison of unprocessed GONG and MDI integrated high frequency (5-7 mHz range) acoustic power maps  {\em (top row)} and  5-7 mHz range egression power maps  {\em (bottom row)} computed using MDI (pupil 15\,-\,60 Mm) and GONG  (pupil 25\,-\,70 Mm) with correction for variable atmospheric smearing applied to GONG  data. 
 \label{fig:20010909_egression}}
\end{figure}

\begin{figure}
\centering
\includegraphics[width=8.2cm]{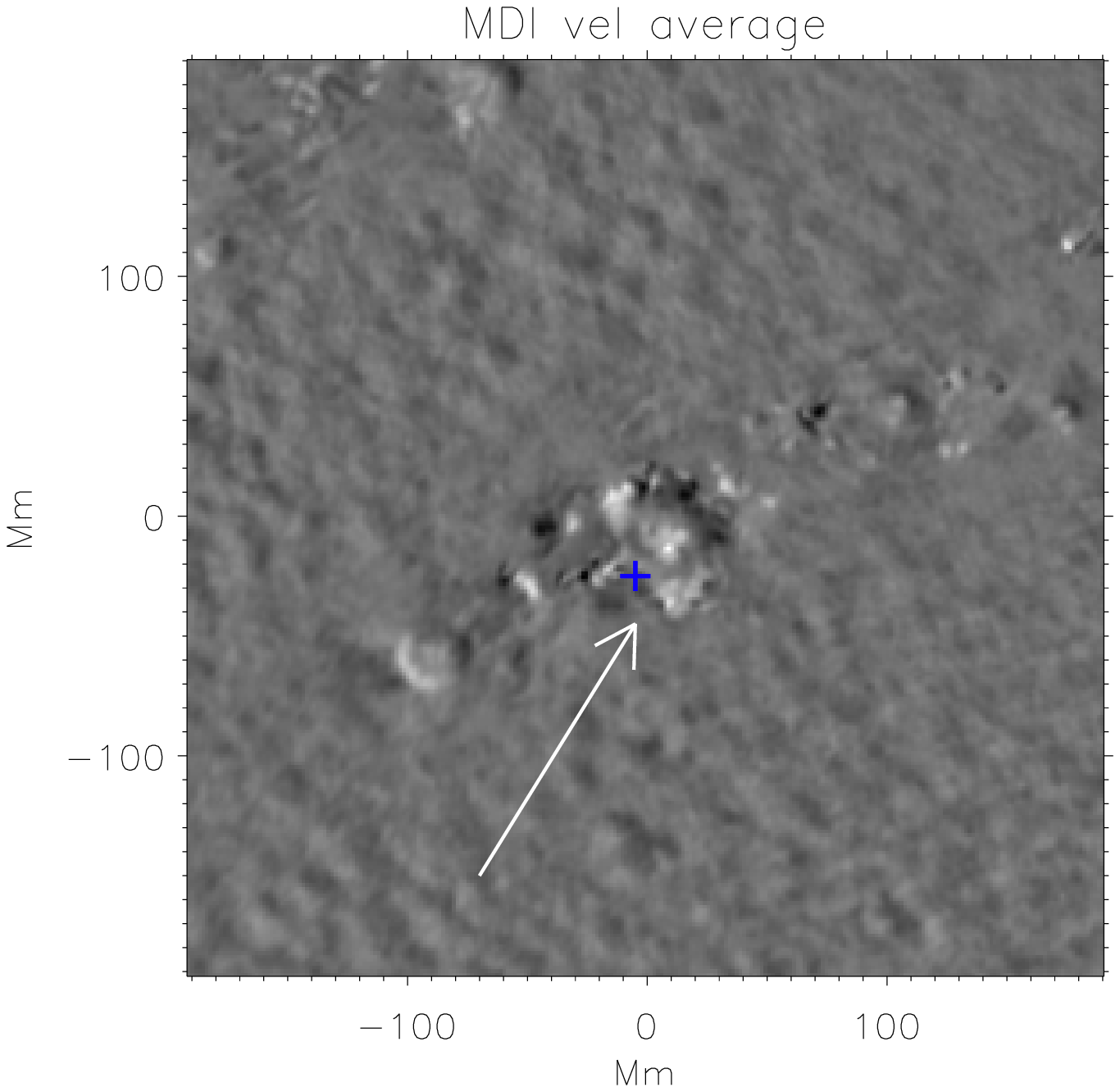}
\includegraphics[width=8.2cm]{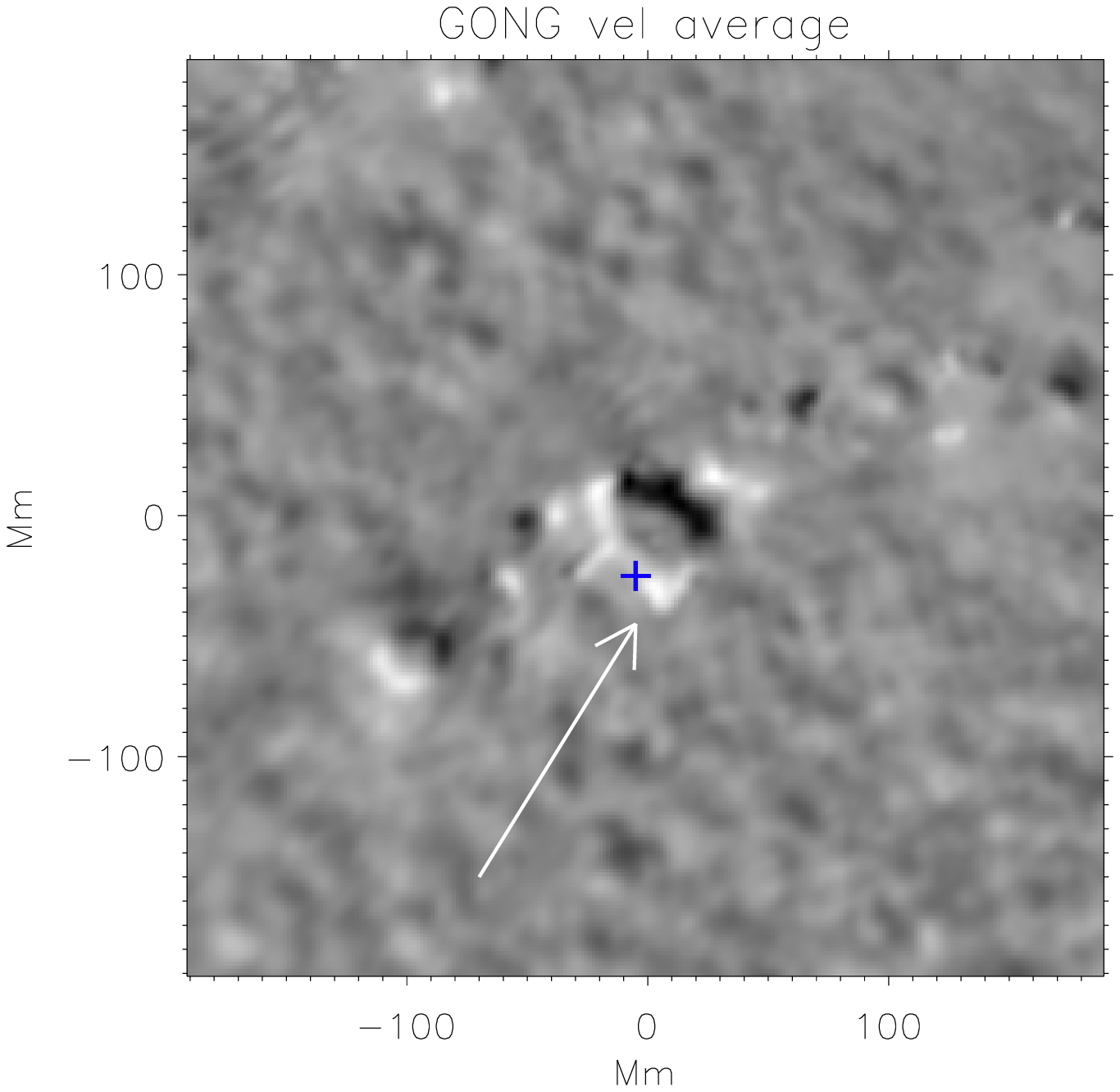}
\includegraphics[width=8.2cm]{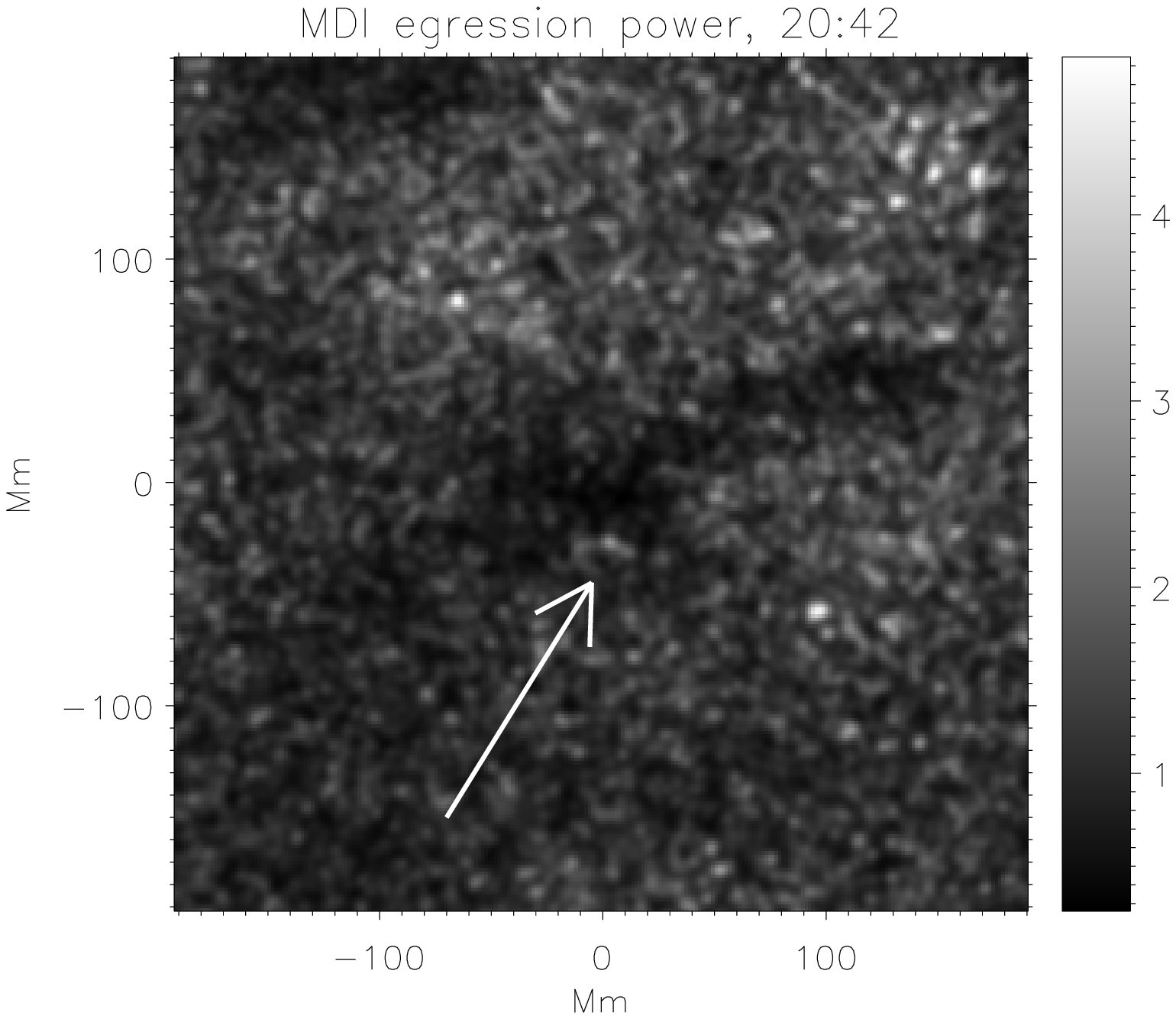} 
\includegraphics[width=8.2cm]{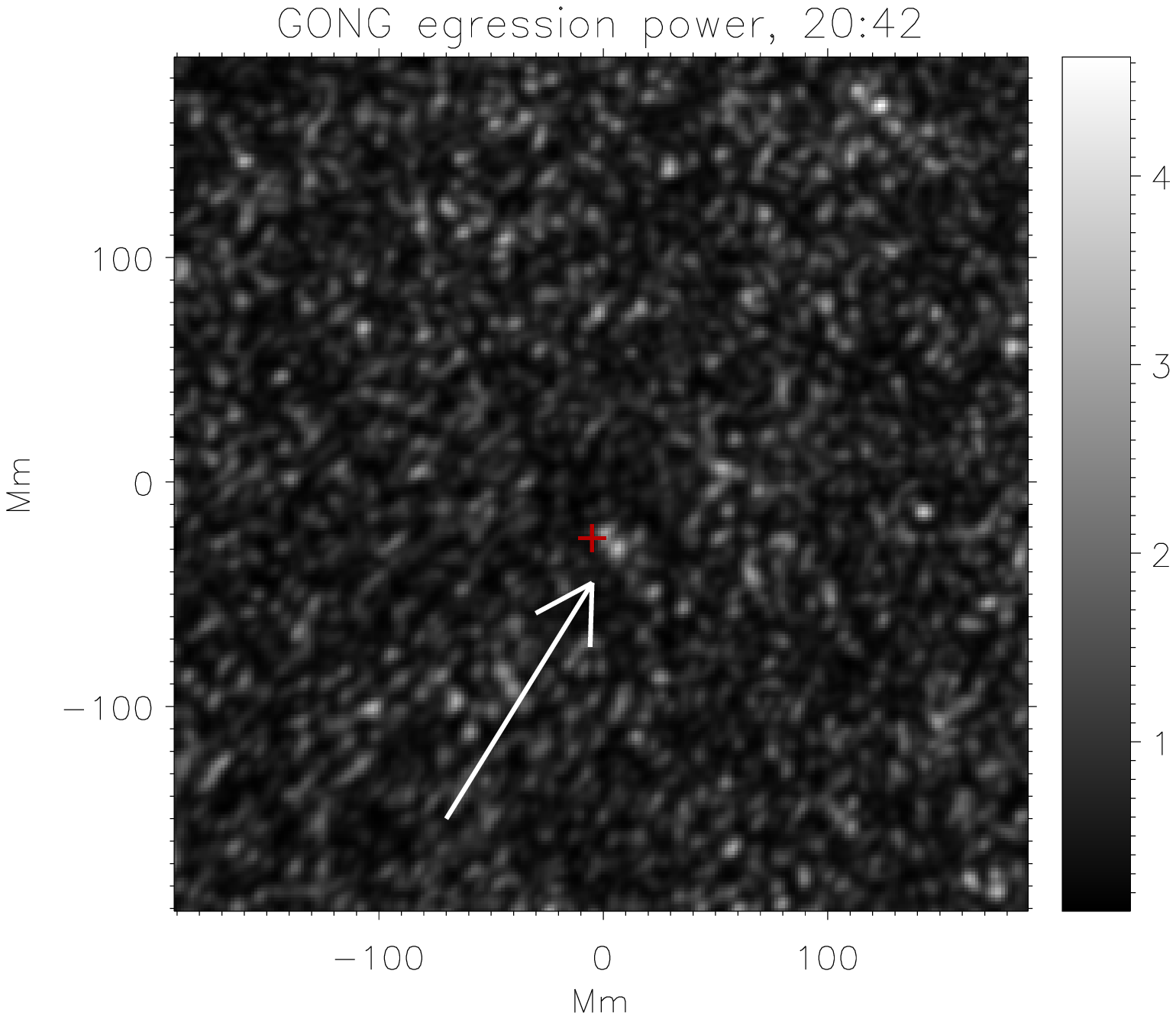}
\includegraphics[width=8.2cm]{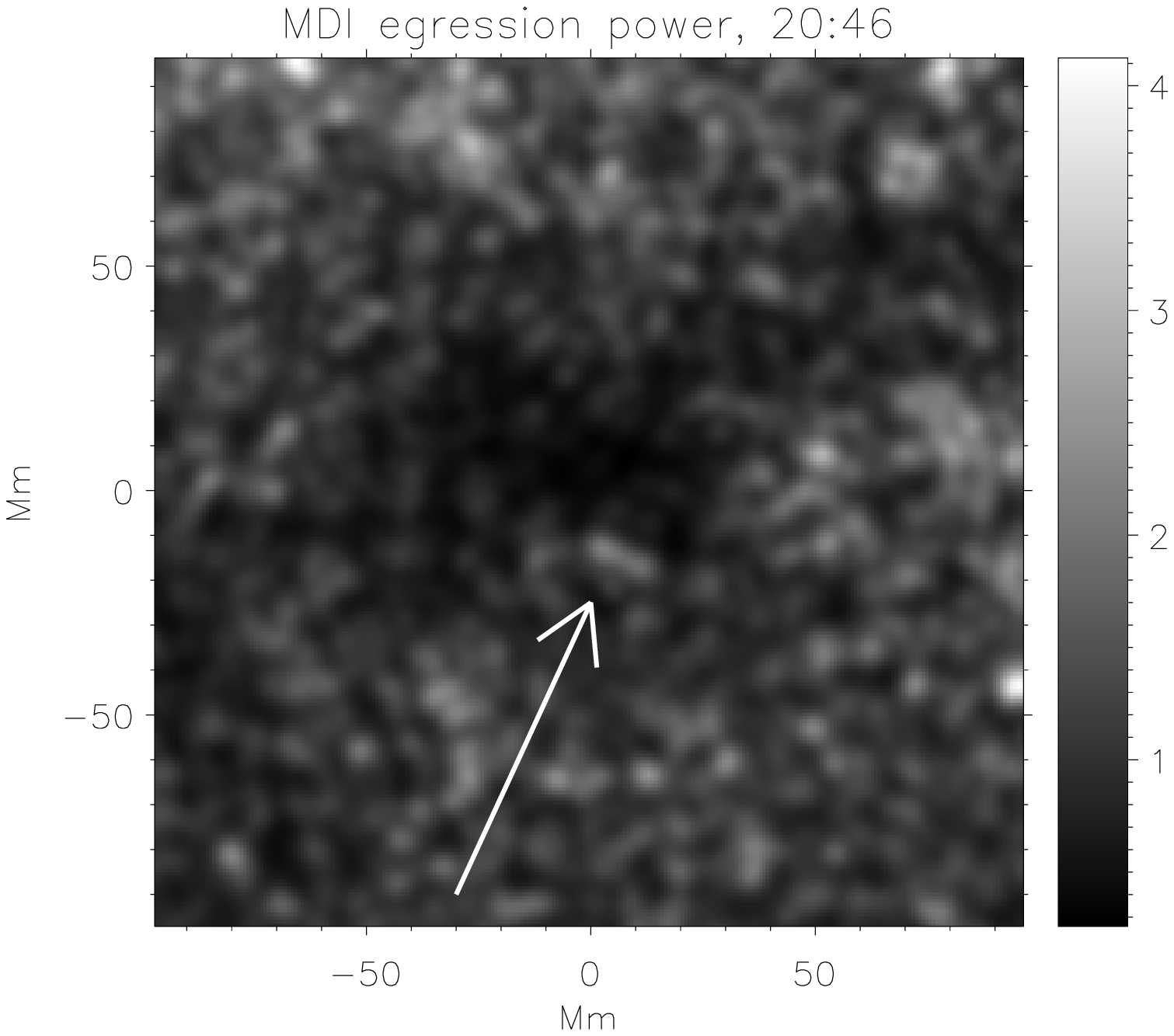}
\includegraphics[width=8.2cm]{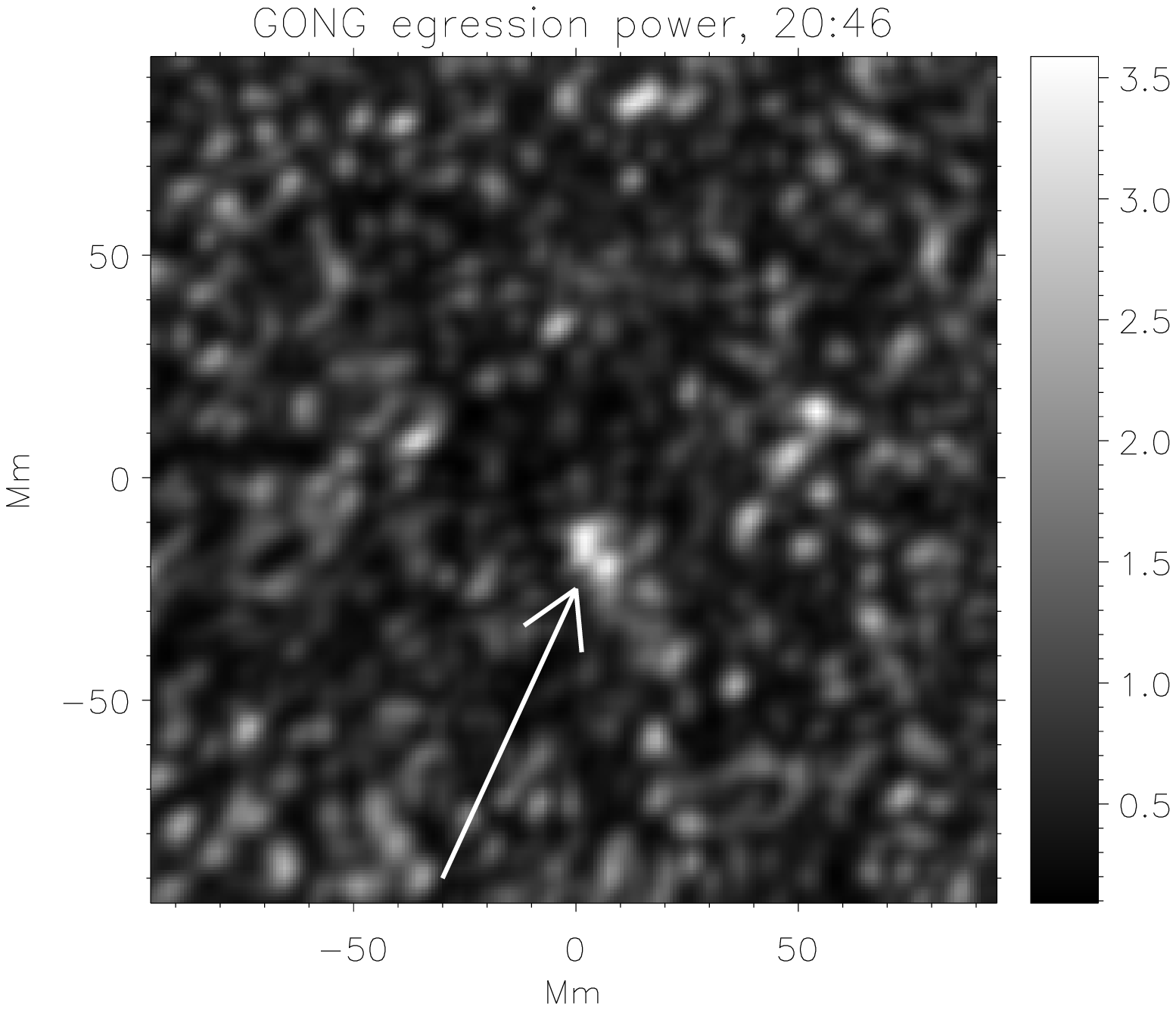} 
\caption{ September 9, 2001 M-class flare: MDI data is in the left column, GONG is on the right. averaged MDI/GONG velocity image (top), followed by egression power snapshots. MDI egression plots reproduce the results in \citet{Donea2006}. Quake location is indicated by an arrow. The plus sign in two upper right frames
indicates the source position assumed for the diagnostics specified by equation (\ref{equ:td}) presented in Figure \ref{fig:20010909_TD}.
 \label{fig:20010909_holo}
}
\end{figure}

\begin{figure}
\centering
\includegraphics[width=8.2cm]{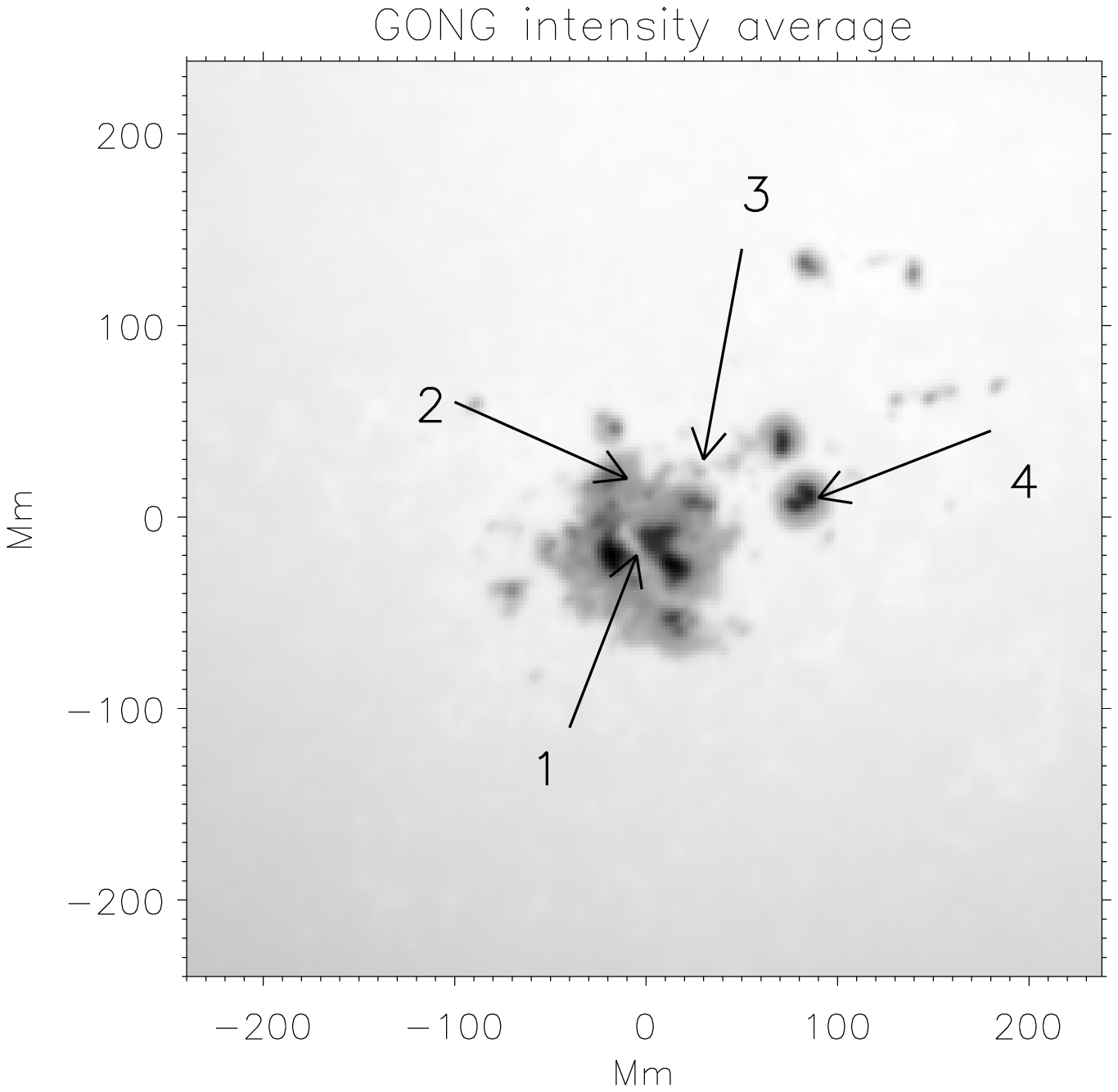}
\includegraphics[width=8.2cm]{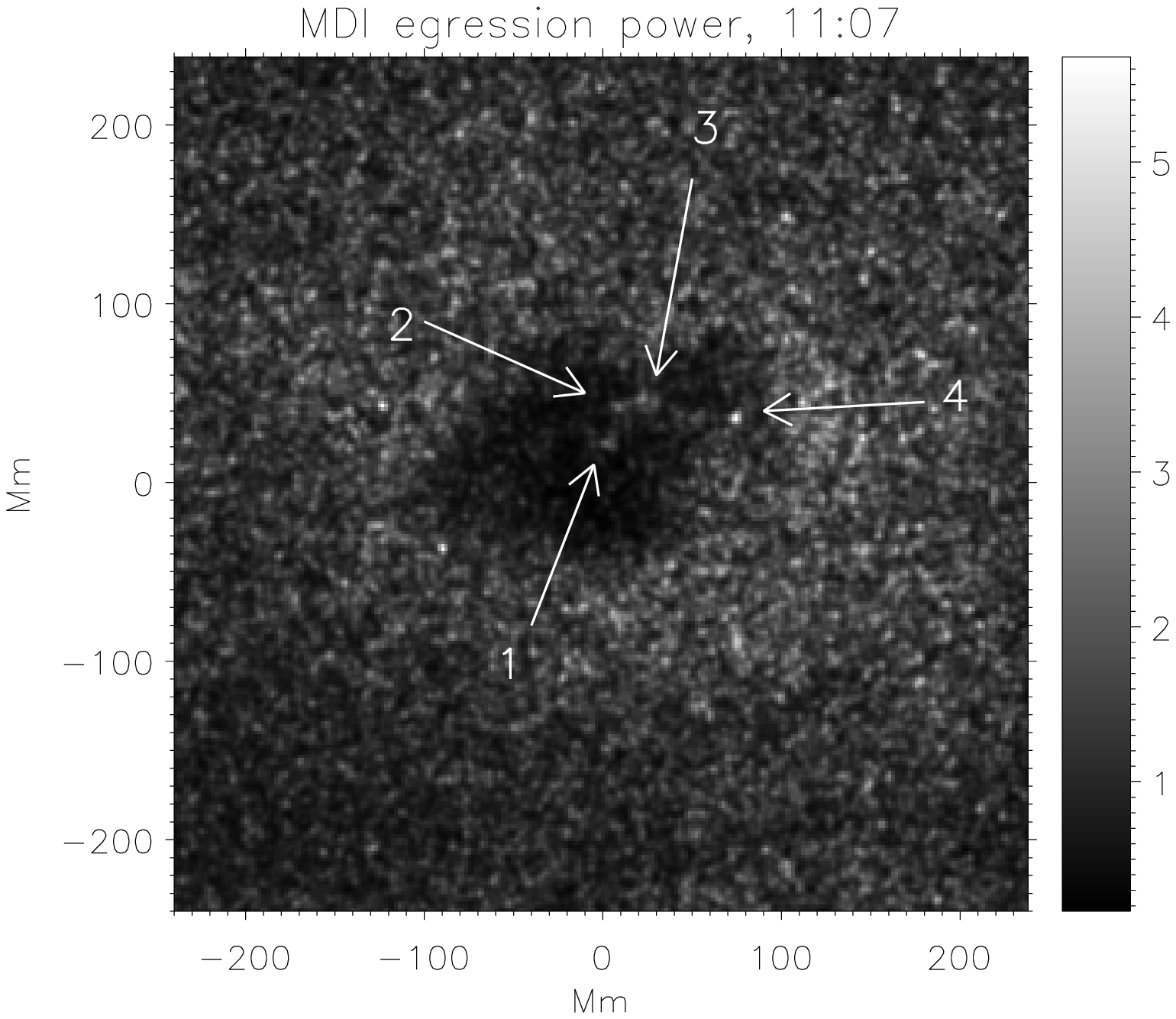}
\includegraphics[width=8.2cm]{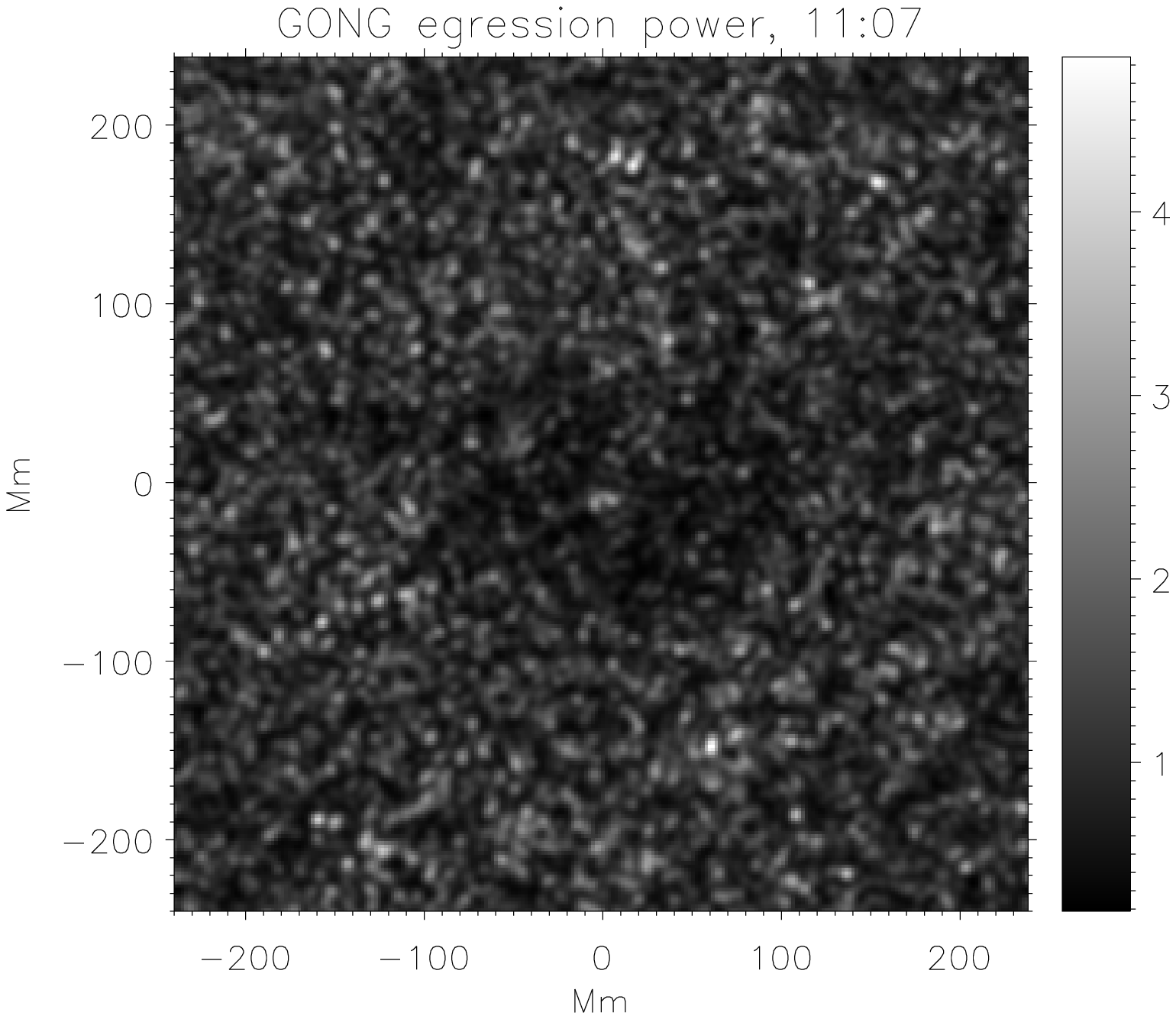}
\includegraphics[width=8.2cm]{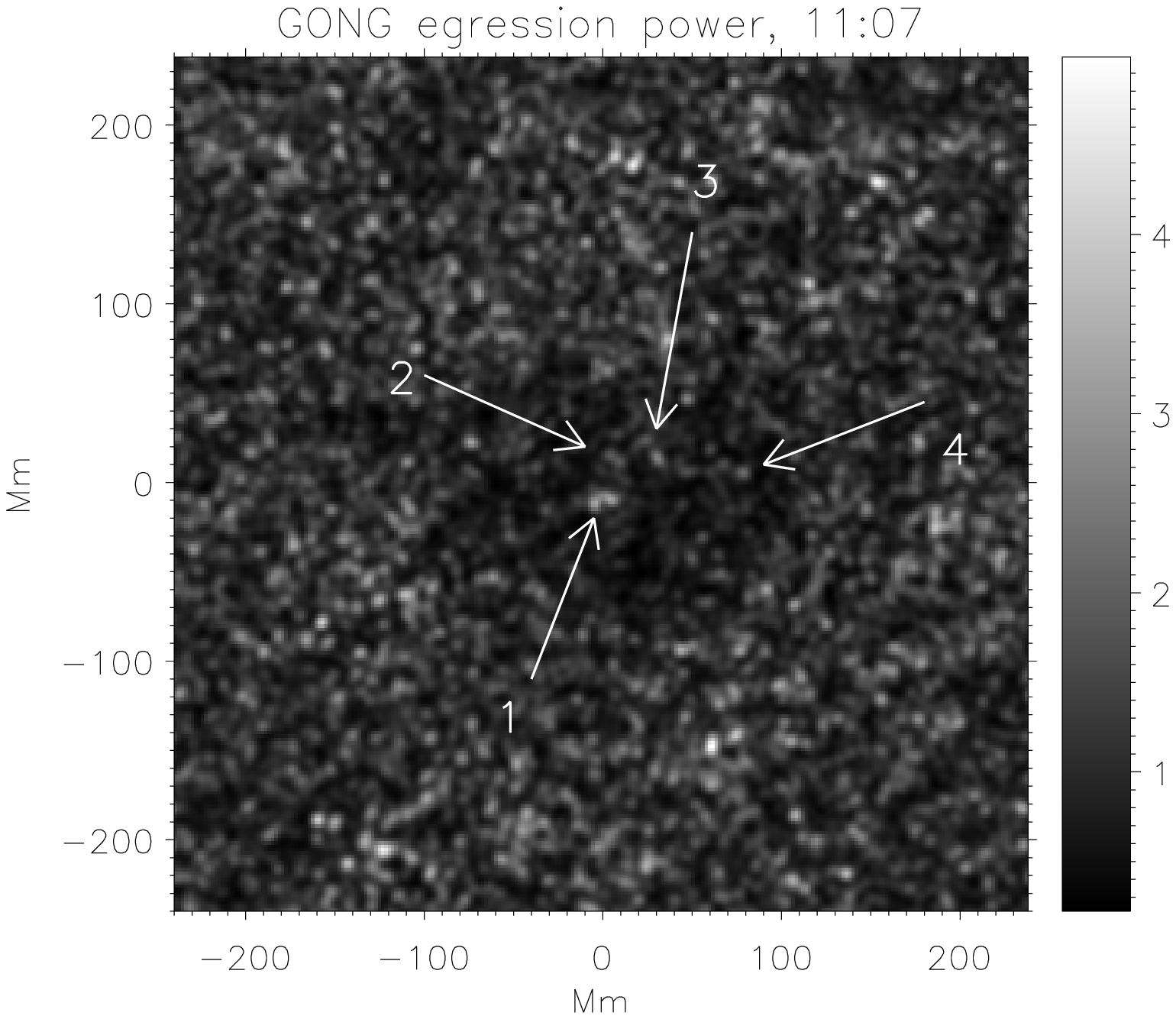} 
\caption{
October 28, 2003 X-class flare: Averaged GONG intensity image is top left, followed by MDI egression snapshot to the right with arrows indicating sources in \citet{DL2005} with \#1, \#3 and \#4 corresponding respectively to time-distance sources \#2, \#1 and \#3 in \citet{Zharkova07}.
At the bottom row: GONG egression power snapshot at 11:07 ({\em left}), and   on the right is the same image with arrows indicating sources in \citet{DL2005} as above. 
 \label{fig:20031028_holo}
}
\end{figure}

\begin{figure}
\centering
\includegraphics[width=8.2cm]{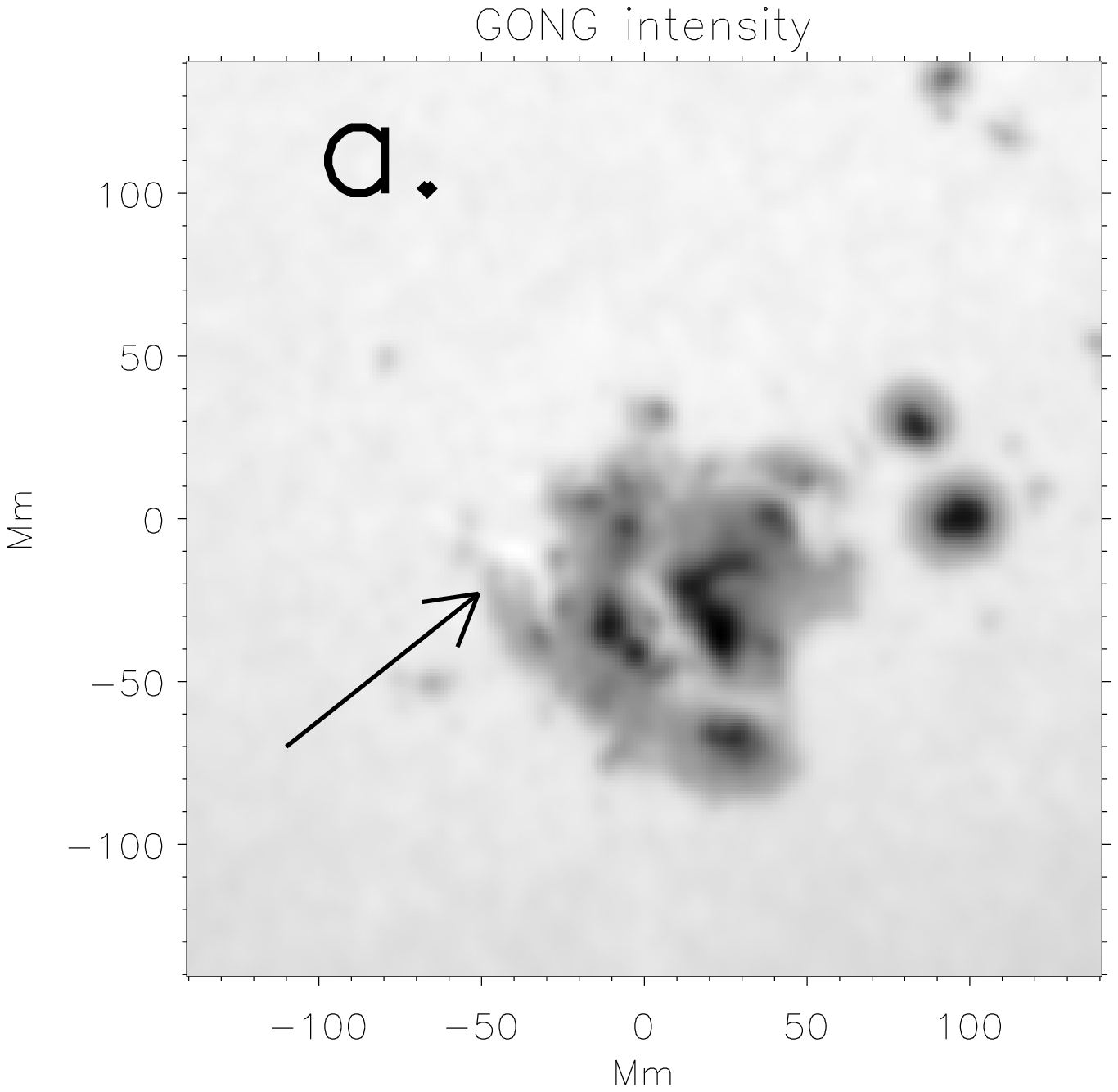} 
\includegraphics[width=8.2cm]{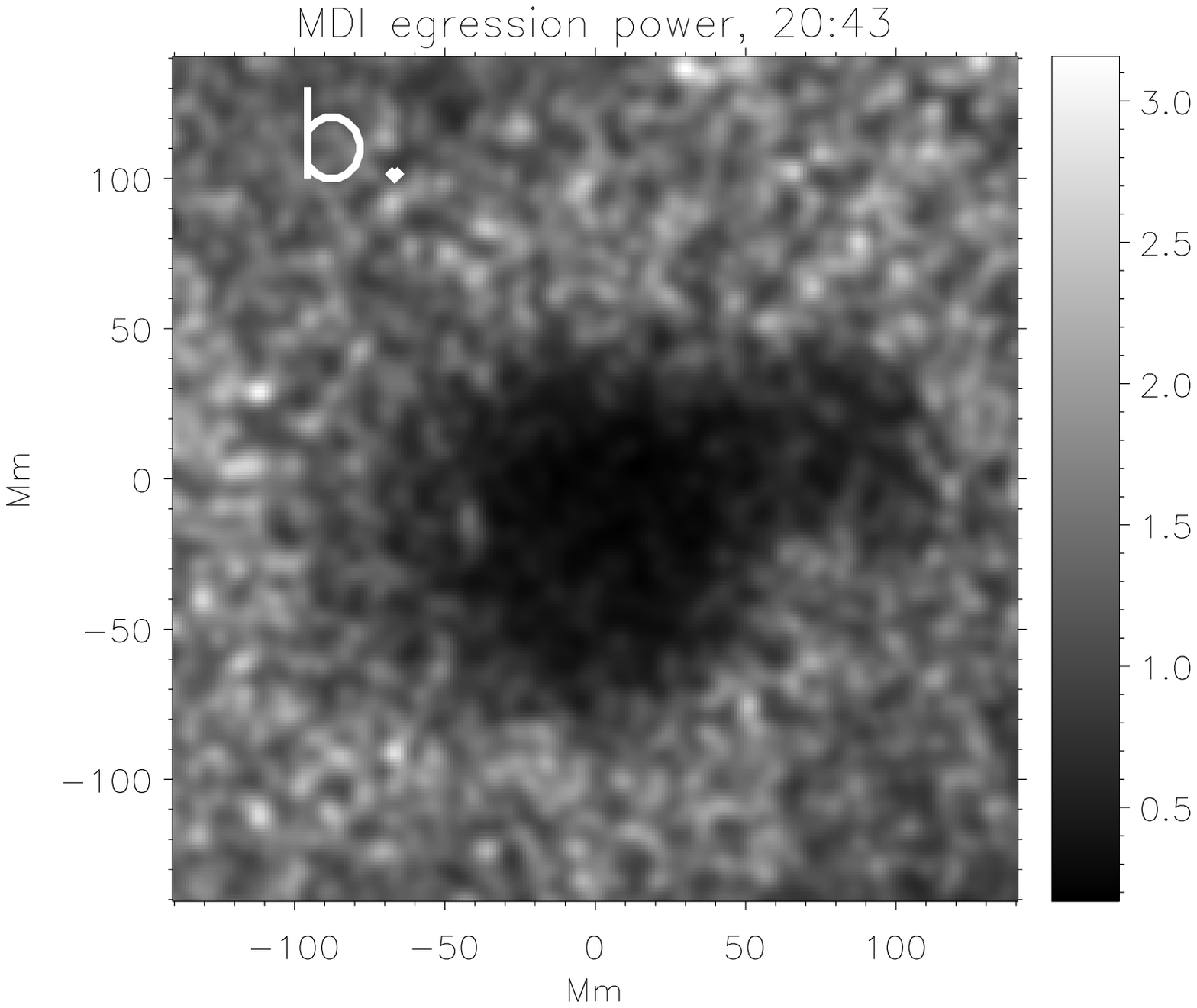}
\includegraphics[width=8.2cm]{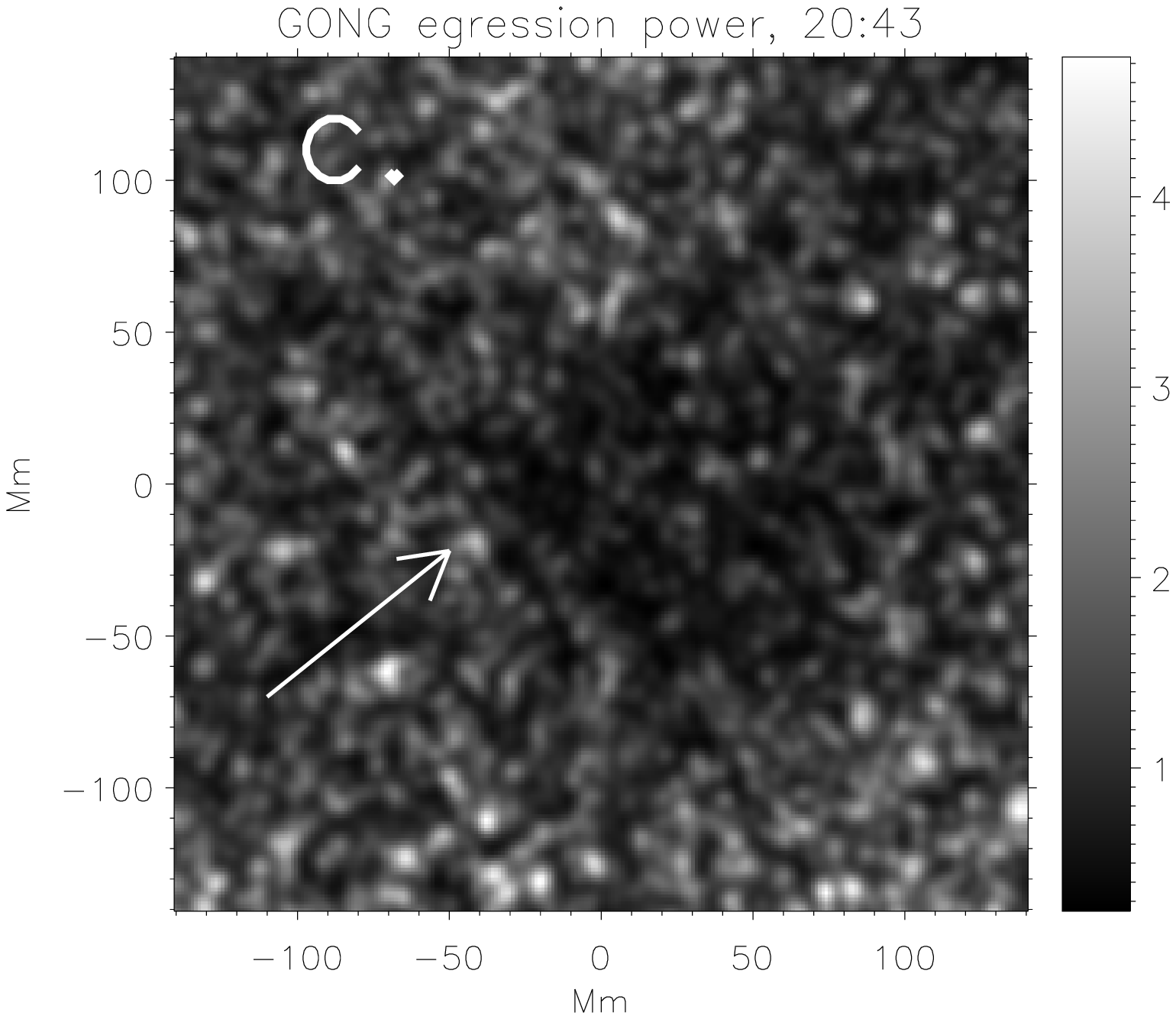}
\includegraphics[width=8.2cm]{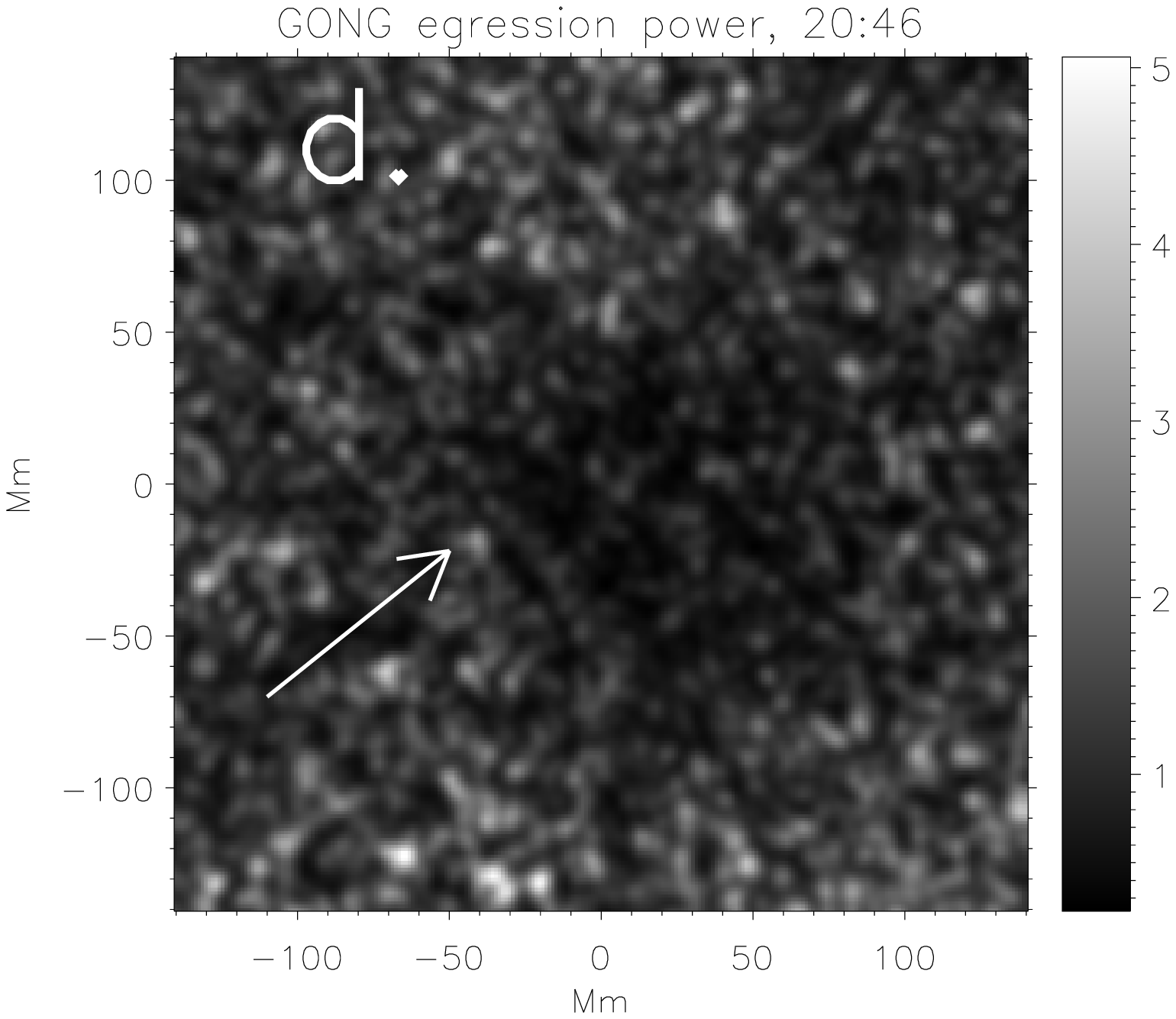}
\includegraphics[width=8.2cm]{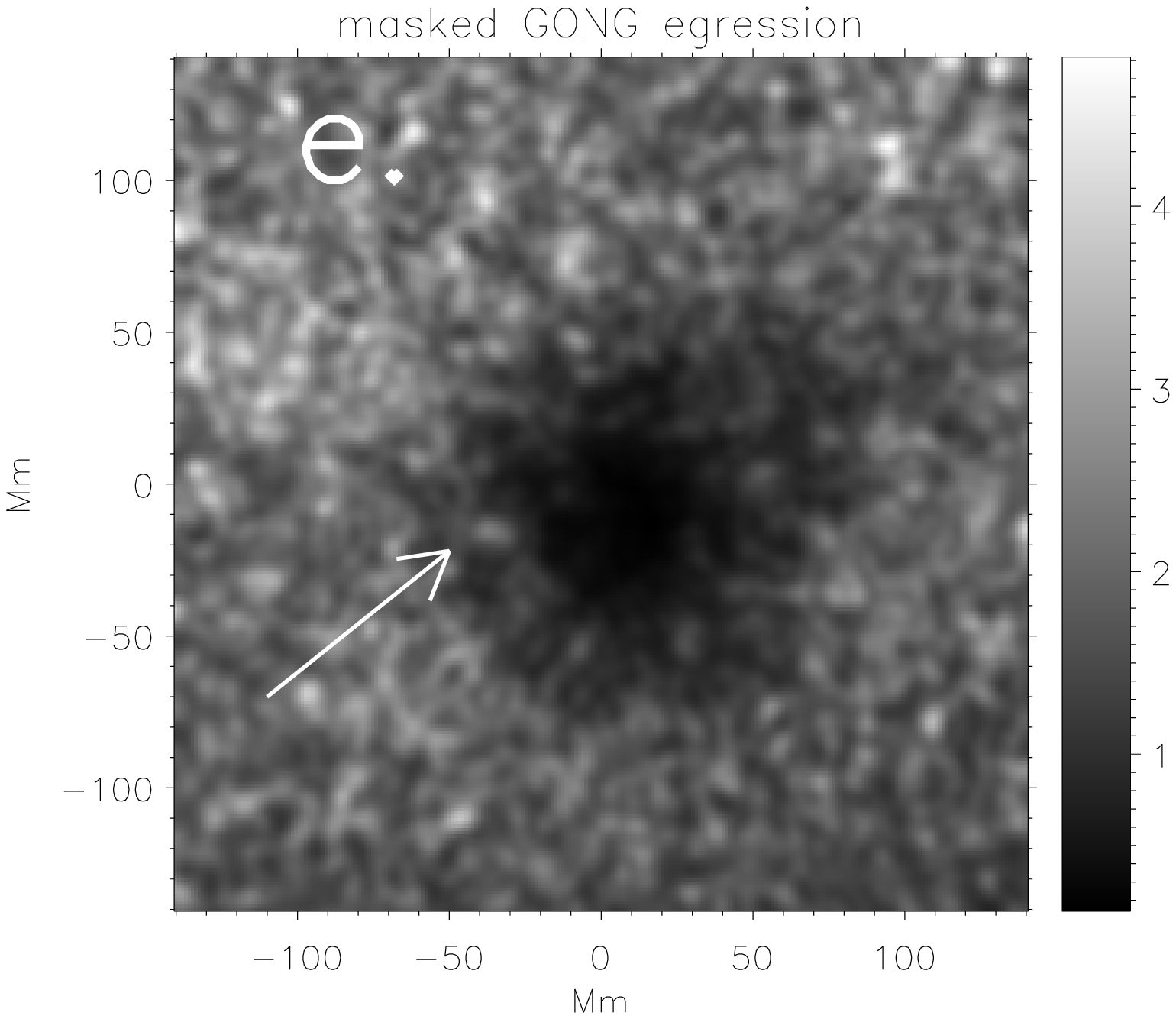}
\includegraphics[width=8.2cm]{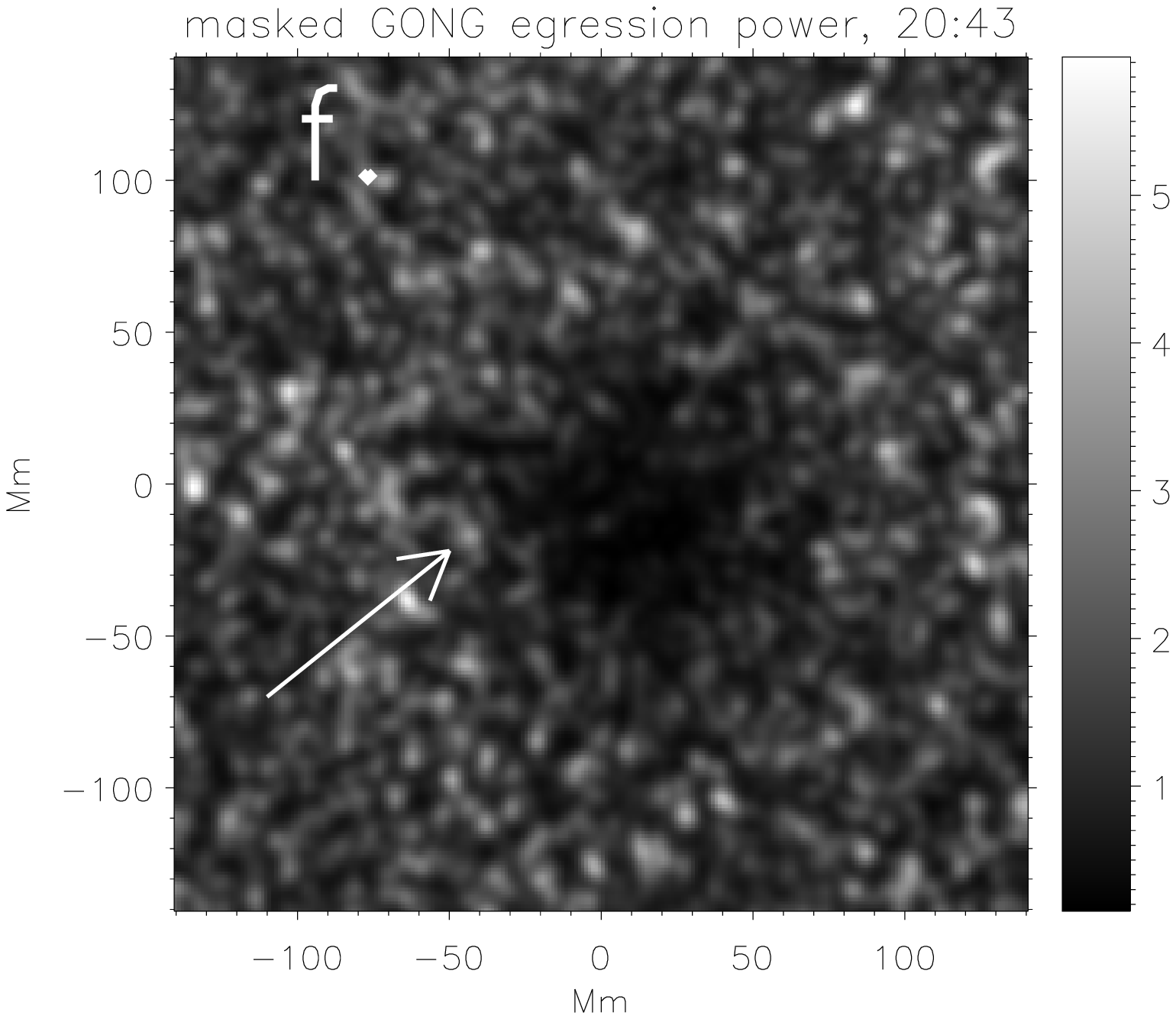}
\caption{October 29, 2003 X-class flare: (a) is GONG intensity image; (b) MDI egression power snapshot at 20:43; (c-d) GONG egression power snapshots around the quake time; (e)  egression power computed from GONG velocity data using sunspot mask  averaged over one hour. Location of the quake is indicated by an arrow; (f) egression power snapshot computed from masked GONG data used in (e) taken around the quake time.
 \label{fig:20031029_holo}}
\end{figure}

\begin{figure}
\includegraphics[width=15.2cm, height=10cm]{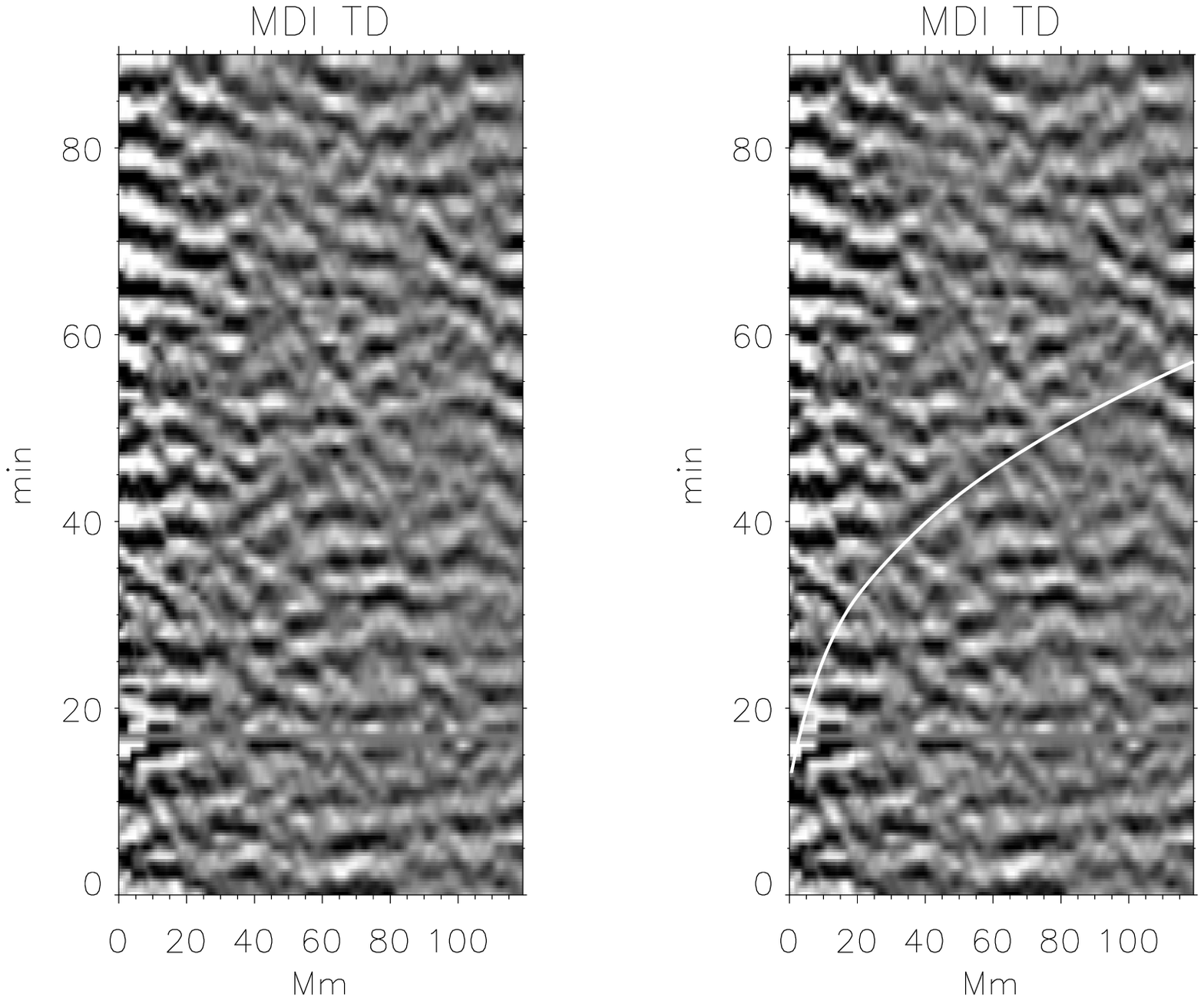} 
\includegraphics[width=15.2cm, height=10cm]{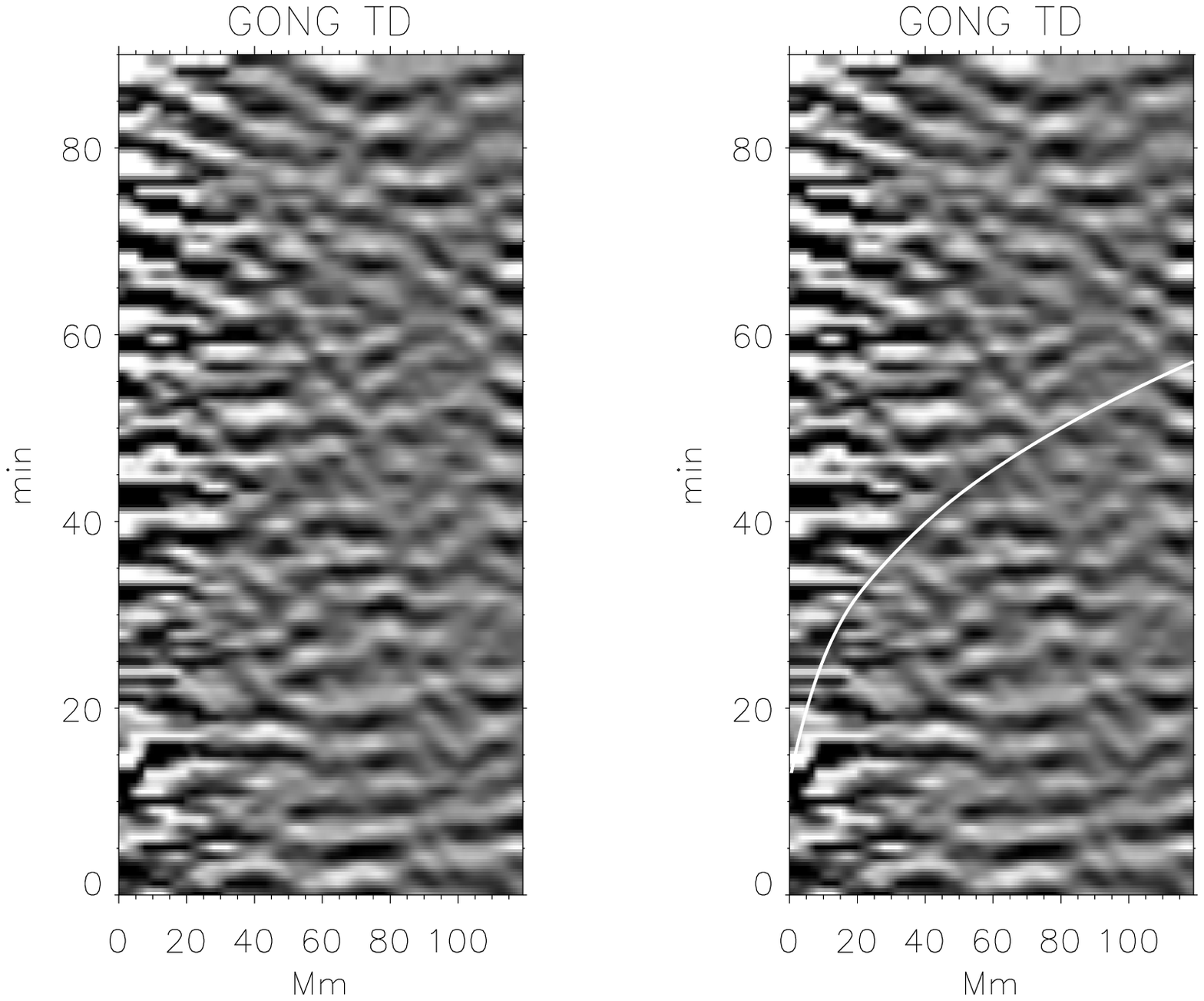}
\caption{September 9, 2001 flare: time-distance diagram computed from MDI velocity data {\em (top row)}, and 
GONG dopplergram observations {\em (bottom)}.
\label{fig:20010909_TD}}
\end{figure}

\begin{figure}
	\centering
\includegraphics[width=15.5cm,  height=9.55cm]{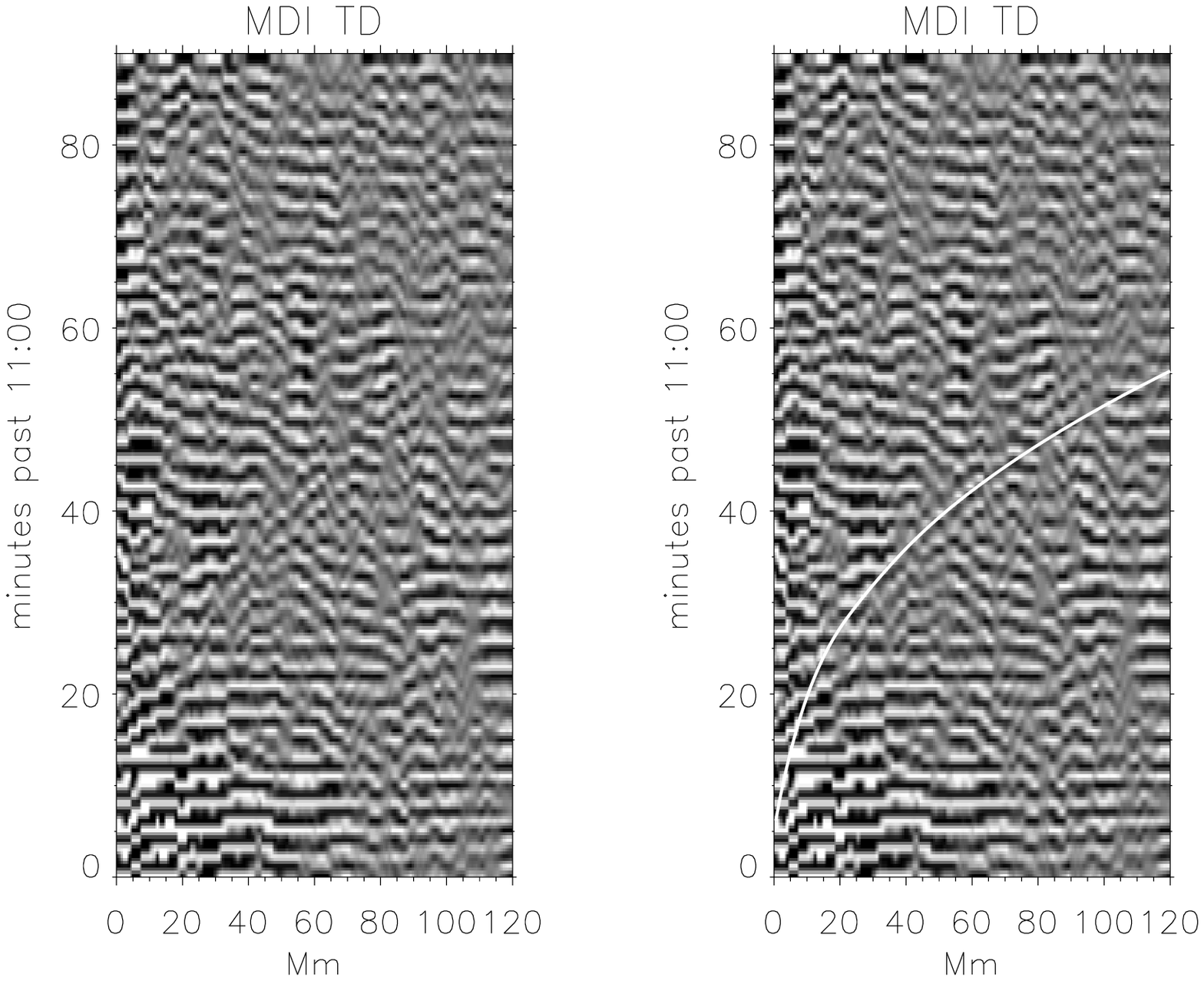}
\includegraphics[width=15.5cm,  height=9.55cm]{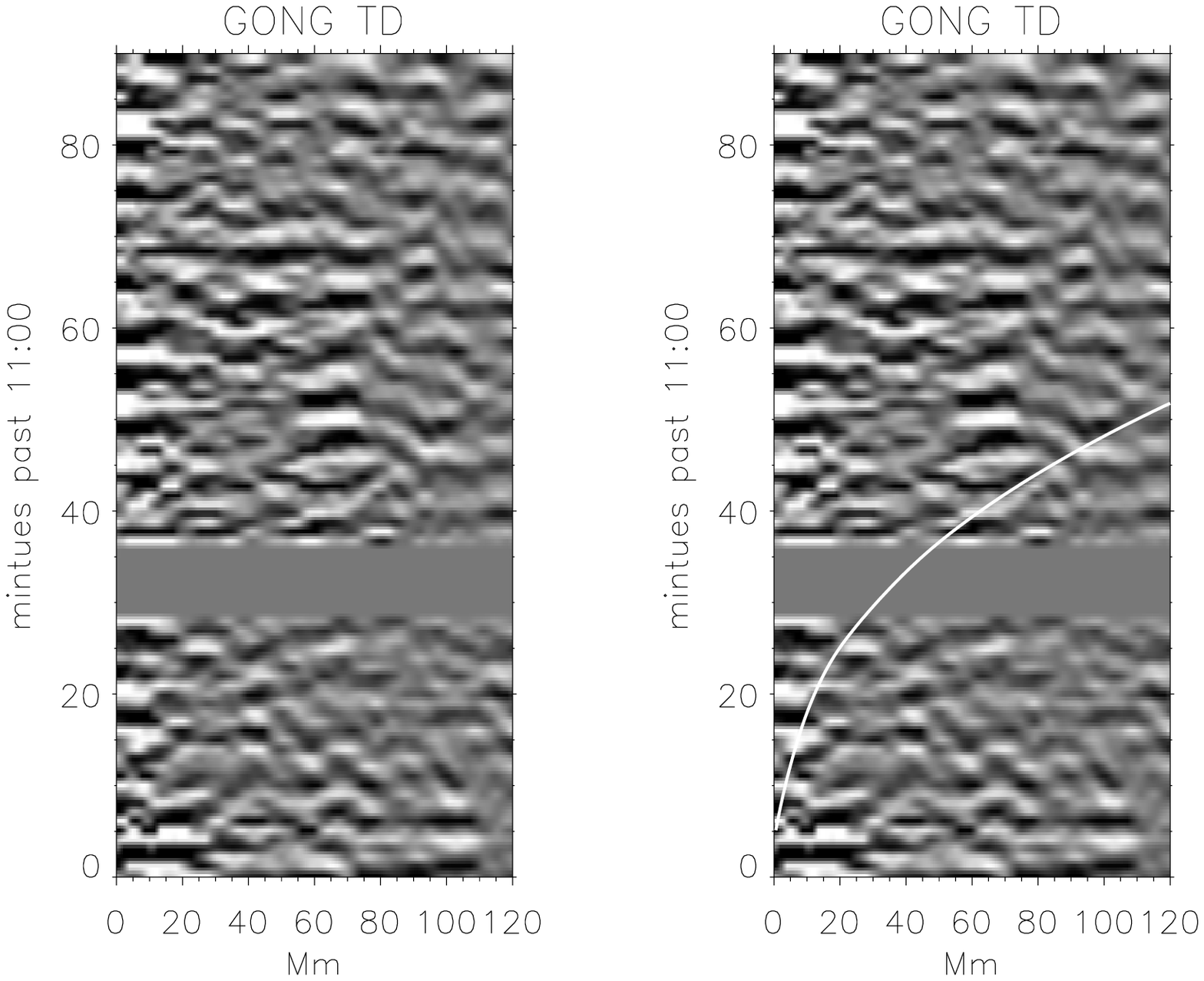}
\caption{October 28, 2003 flare: {\em (Top row)}:time-distance diagram computed from MDI data. The plots are reproduced from \citet{Zharkova07}.  {\em (Bottom row)}: time-distance diagram computed from GONG data. 0 along the $y$-axis corresponds to 11:00 UT.
\label{fig:20031028_TD}}
\end{figure}

\bibliographystyle{aa}

\end{document}